\documentclass[12pt]{article}
\usepackage{amsbsy,amsmath}
\usepackage{amsfonts}
\usepackage{amssymb}
\usepackage{amscd}
\usepackage{bbm}
\usepackage{fancybox}
\usepackage{cite}
\usepackage{amsmath,amsfonts,amsbsy}
\usepackage{pstricks,pst-node}
\usepackage[small,bf,hang]{caption2}
\usepackage{graphicx}
\usepackage{epsfig}
\usepackage{psfrag}
\usepackage{comment}

\usepackage[mathscr]{euscript}

\usepackage{float}

\psset{unit=1.3cm,linewidth=.5pt,radius=.2}  % controls the size of diagrams

\usepackage{multirow}                     % tables cells spanning multiple rows
\usepackage{float}                          % to force floaters to stay Here
\usepackage{lscape}                         % landscape laid out pages
\usepackage{bm}

\newcommand{\be}{\begin{equation}}
\newcommand{\ee}{\end{equation}}
\newcommand{\bea}{\begin{eqnarray}}
\newcommand{\eea}{\end{eqnarray}}
\newcommand{\ba}{\begin{array}}
\newcommand{\ea}{\end{array}}

\def\HH{\mathcal{H}}
\def\LL{\mathcal{L}}
\def\hh{\mathfrak{h}}
\def\TT{\mathcal{T}}

\font\mybb=msbm10 at 12pt
\def\bb#1{\hbox{\mybb#1}}

\makeatletter \@addtoreset{equation}{section} \makeatother

\def\slashchar#1{\setbox0=\hbox{$#1$}           % set a box for #1
   \dimen0=\wd0                                 % and get its size
   \setbox1=\hbox{/} \dimen1=\wd1               % get size of /
   \ifdim\dimen0>\dimen1                        % #1 is bigger
      \rlap{\hbox to \dimen0{\hfil/\hfil}}      % so center / in box
      #1                                        % and print #1
   \else                                        % / is bigger
      \rlap{\hbox to \dimen1{\hfil$#1$\hfil}}   % so center #1
      /                                         % and print /
   \fi}

\begin{document}

\begin{titlepage}%1

\begin{center}

\vskip 1.5cm

{\Large \bf Causal  chiral 2-form electrodynamics}

\vskip 1cm

{\bf Jorge G.~Russo\,${}^{a,b}$ and  Paul K.~Townsend\,${}^c$} \\

\vskip 25pt

{\em $^a$  \hskip -.1truecm
\em Instituci\'o Catalana de Recerca i Estudis Avan\c{c}ats (ICREA),\\
Pg. Lluis Companys, 23, 08010 Barcelona, Spain.
 \vskip 5pt }

\vskip .4truecm

{\em $^b$  \hskip -.1truecm
\em Departament de F\' \i sica Cu\' antica i Astrof\'\i sica and Institut de Ci\`encies del Cosmos,\\ 
Universitat de Barcelona, Mart\'i Franqu\`es, 1, 08028
Barcelona, Spain.
 \vskip 5pt }
 
 \vskip .4truecm

{\em $^c$ \hskip -.1truecm
\em  Department of Applied Mathematics and Theoretical Physics,\\ Centre for Mathematical Sciences, University of Cambridge,\\
Wilberforce Road, Cambridge, CB3 0WA, U.K.\vskip 5pt }

\hskip 1cm

\noindent {\it e-mail:}  {\texttt jorge.russo@icrea.cat, pkt10@cam.ac.uk}

\end{center}

\vskip 0.5cm
\begin{center} {\bf ABSTRACT}\\[3ex]
\end{center}

Generic nonlinear theories of chiral 2-form electrodynamics allow superluminal 
propagation in some stationary homogeneous backgrounds and are therefore acausal. We 
find a simple parameterisation of the Hamiltonian for causal theories, and we use it to 
show that the stress-energy tensor satisfies both the Dominant and Strong energy conditions. 
We also revisit the Perry-Schwarz formulation, clarifying aspects of it and of its relation to the 
Hamiltonian formulation.

\noindent

\vskip 1cm

{\large {\it In Memoriam}: Alexandru Luca Mezincescu}

\vfill

\end{titlepage}
\tableofcontents
%%%%%%%%%%%%%%%%%%%%

%%%%%%%%%%%%%%%%%%%%%
\section{Introduction}
\setcounter{equation}{0}

The free-field equations of six-dimensional (6D) chiral 2-form electrodynamics are equivalent to 
a self-duality condition on the 3-form field-strength of a 2-form gauge potential. These 6D field equations 
are conformal, like the 4D Maxwell equations, but with plane-wave solutions that have three 
(rather than two) independent polarisations. One reason for interest in chiral 2-form electrodynamics 
is that it has superconformal extensions for both $(1,0)$ and $(2,0)$ 6D-supersymmetry \cite{Howe:1983fr}, and another 
is that the low-energy effective action for the M5-brane is closely related to a particular {\sl nonlinear} $(2,0)$-supersymmetric theory of chiral 2-form electrodynamics \cite{Howe:1996yn,Perry:1996mk,Pasti:1997gx,Howe:1997fb,Howe:1997vn,Bandos:1997ui}.

Manifestly Lorentz-invariant Lagrangian densities for chiral 2-form electrodynamics theories necessarily include fields other than 
the 3-form field-strength and have additional gauge-invariances; e.g. \cite{Siegel:1983es,Pasti:1996vs,Avetisyan:2022zza}. A simple 
alternative is the Hamiltonian formulation, with field equations obtainable from the  manifestly $SO(5)$-rotation 
invariant  ``phase-space action''
\cite{Henneaux:1988gg,Bandos:2020hgy}
\be\label{6Dact}
I[\bb{A}]= \int\! dt\int\! d^5x \left\{  \tfrac12 \dot{\bb{A}} \cdot \bb{B} - \mathcal{H}(\bb{B})\right\} \, , 
\ee
where $\bb{A}$ is an antisymmetric tensor on Euclidean 5-space, with components $A_{ij}$, and 
$\bb{B}$ is its ``5-space curl'',
\be
B^{ij} =  (\boldsymbol{\nabla}\times \bb{A})^{ij} := \tfrac12 \varepsilon^{ijklm} \partial_k A_{lm} \, , 
\ee
which is invariant under  the gauge transformation $A_{ij} \to A_{ij} + 2\partial_{[i}\alpha_{j]}$.
The inner product of any two antisymmetric tensors $\bb{C}$ and $\bb{C}'$ is defined as 
\be
\bb{C} \cdot\bb{C}' := \tfrac 12 C^{ij} C'_{ij} \qquad \big(\Rightarrow \ |\bb{C}|^2 = \tfrac12 C^{ij}C_{ij}\big),  
\ee
and the action is therefore gauge-invariant up to a boundary term because of the Bianchi identity 
$\partial_iB^{ij} \equiv 0$.  It is an action for {\sl chiral} 2-form electrodynamics because the antisymmetric tensor density canonically conjugate to $\bb{A}$ has been identified with $\bb{B}$. This identification eliminates the need to impose a Gauss-law type constraint via a Lagrange multiplier field $A_{i0}$, and it
implies that the rotation-invariant Hamiltonian density $\HH$ is a function of $\bb{B}$ only.  The free-field case is $\HH =  \frac12 |\bb{B}|^2$. 

All rotation-invariant functions of $\bb{B}$ can be expressed as functions of the two variables \cite{Bandos:2020hgy}
\be\label{6Dsp}
s= \frac12 |\bb{B}|^2 \, , \qquad p= |\bb{B} \times \bb{B}| \, , 
\ee
where $p$ is the magnitude of the field momentum density\footnote{The sign is opposite for chiral and anti-chiral theories; the choice made here is convenient when considering dimensional reduction to 4D.} ${\bf p} = -\bb{B} \times \bb{B}$; the cross product is defined by
\be
[\bb{B} \times \bb{B}]_i :=  \frac18 \varepsilon_{ijklm} B^{jk}B^{lm}\, . 
\ee
The requirements of spacetime translation invariance and  rotation invariance imply that $\HH$ is expressible as a function of 
$(s,p)$ only. Another useful basis is\footnote{This differs from the $(u,v)$ basis in 
\cite{Bandos:2020hgy} by an interchange of $u$ with $v$.}
\be\label{uvdefs}
u= \frac12 \left(s- \sqrt{s^2-p^2}\right) \, , \qquad v = \frac12 \left(s+ \sqrt{s^2-p^2}\right)\, . 
\ee
Notice that $v\ge u\ge0$ if we adopt the convention that the square-roots 
are positive rather than negative, but a curve in the $(u,v)$-plane  that reaches its $v=u$ boundary could be smoothly continued into the region $u\ge v\ge0$ by viewing it as a second branch of the functions $u$ and $v$ in terms of $s$ and $p$. The status of the $v=u$ boundary of the region $v\ge u\ge0$ is therefore different from the status of its $u=0$ boundary. In principle, 
we should consider the entire positive quadrant of the $(u,v)$ plane as the largest possible domain of $\HH(u,v)$, with two branches that meet on the line $v=u$. However, for most purposes here it will be sufficient to consider only the region $v\ge u\ge 0$, which we shall refer to as the ``physical region'' of the $(u,v)$ plane. 

The action of \eqref{6Dact} is not obviously invariant under 6D Lorentz boosts, and the field equations are not Lorentz invariant for generic 
$\HH(u,v)$. 
The condition on this function for Lorentz invariant field equations is \cite{Bandos:2020hgy}
\be\label{HamLI}
\HH_u\HH_v=1\, . 
\ee
This equation also arises in the Hamiltonian formulation of  nonlinear self-dual electrodynamics (NLED) as a condition for 4D Lorentz invariance, with a different interpretation of the variables $(u,v)$ but again defined for $v\ge u\ge0$ \cite{Bialynicki-Birula:1984daz,Bandos:2020jsw}.  This implies a one-to-one correspondence between the 4D and 6D interpretations of solutions of \eqref{HamLI}, which are related by a dimensional reduction/truncation procedure that we review in subsection \ref{subsec1}. Inspired by work of Perry and Schwarz that we mention below, and using results from Courant and Hilbert \cite{C&H}, we showed in \cite{Russo:2024ptw} that the general solution of \eqref{HamLI} for boundary condition $\HH(0,v) = \hh(v)$ is 
\be\label{CH-Ham}
\HH(u,v) = \hh(\sigma) + \frac{2u}{\hh^\prime(\sigma)} \, , \qquad \sigma = v- \frac{u}{[\hh^\prime(\sigma)]^2}\, , 
\qquad  \hh^\prime(\sigma) >0\, .
\ee
It is straightforward to verify that this implies
\be\label{firstds}
\HH_v = \hh^\prime\, , \qquad \HH_u = 1/\hh^\prime\, ,
\ee
and hence \eqref{HamLI}.  For $\hh(v) \equiv 0$, there is an additional solution: $\HH = \sqrt{4uv} = p$. This defines the conformal but interacting Bialynicki-Birula (BB) electrodynamics in the 4D case \cite{Bialynicki-Birula:1984daz,Bialynicki-Birula:1992rcm} and its generalisation to chiral 2-form electrodynamics in the 6D case \cite{Gibbons:2000ck,Townsend:2019ils}. All other solutions of \eqref{CH-Ham} also have both a 4D and a 6D interpretation. For example
\be\label{hBI}
\hh(\sigma) = \sqrt{T^2+2T\sigma}-T \quad \Rightarrow \quad \HH(u,v) = \sqrt{(T+2u)(T+2v)} -T\, , 
\ee
which is the Born-Infeld theory for 4D and the `M5' chiral 2-form electrodynamics for 6D. More generally, {\sl we assume that $h(\sigma)$ has a Taylor expansion about $\sigma=0$} since $\HH$ then has a weak-field expansion.  
Omitting a constant term in $\hh$ since it contributes only to the vacuum energy, and taking into account that $\hh'>0$,  we have 
\be\label{Taylor}
\hh(\sigma) = e^{-\gamma} \sigma + \mathcal{O}(\sigma^2)\, ,  
\ee
for some constant $\gamma$. By itself, the linear term in $\sigma$ yields the Hamiltonian density for the general conformal weak-field limit. This is Maxwell for 
$\gamma=0$ and ModMax for $\gamma>0$  \cite{Bandos:2020jsw} (and an acausal variant of it for $\gamma<0$). 

The above construction provides a parametrisation of the Hamiltonian densities of both 6D chiral 2-form electrodynamics 
and self-dual 4D NLED  in terms of the one-variable function $\hh$, which we call a Courant-Hilbert (CH) function.  For the 4D interpretation of $\HH(u,v)$ there is also a correspondence (via Legendre transform) with a 4D manifestly Lorentz-invariant  Lagrangian density that can be written in terms of a Lagrangian CH-function $\ell$. Conditions on $\ell$ that are both necessary and sufficient for causality were  found in \cite{Russo:2024llm}  and were `translated' in  \cite{Russo:2024ptw} to conditions on $\hh$, with the following result: 
\be\label{causal-h1}
\hh^\prime (\sigma) \le1\, , \qquad \hh^\prime{}^\prime(\sigma) \le0\, , 
\ee
and 
\be\label{causal-h2}
\hh^\prime(\sigma) + 2\sigma \hh^\prime{}^\prime >0\, . 
\ee
Because the 4D Hamiltonian and Lagrangian densities are Legendre dual,  these causality constraints on $\hh(\sigma)$ come with  
the following restriction on its domain:
\be\label{zero-sig}
\sigma\ge0\, .
\ee
Since each choice of $(u,v)$ defining a  background solution is associated with a particular value of $\sigma$, this restriction is  potentially
a restriction on the physical domain of $\HH(u,v)$; i.e. the values of $(u,v)$ corresponding to possible field configurations. As we shall see, there
is no such restriction if the weak-field limit is Maxwell ($\gamma=0$) but  if  the weak-field limit is ModMax then $\HH$ (viewed as a function of the Legendre dual of the electric field) is restricted to its convex domain, as found for ModMax itself in \cite{Bandos:2020jsw}.

In the 6D context we should expect the inequalities of  \eqref{causal-h1} and \eqref{causal-h2} to be  necessary (at least) for the causal propagation of plane wave perturbations of backgrounds with $\sigma\ge0$. This is because superluminal waves in a 4D background `lift' to superluminal waves in a corresponding 6D background, for reasons that we explain in subsection \ref{subsec1}. What is less clear is whether we are entitled to assume the condition \eqref{zero-sig} in the 6D context, because it is not enforced in any obvious way by some Hamiltonian-Lagrangian equivalence.

There is a potential  equivalence of the Hamiltonian formulation to the Perry-Schwarz (PS) formulation \cite{Perry:1996mk} because they are gauged-fixed versions of, respectively,  the ``timelike'' and ``spacelike''  PST formulations \cite{Bandos:2020hgy} and there is evidence that the PST field equations are independent of the timelike/spacelike choice \cite{Ferko:2024zth}, but it is not clear to us how this equivalence would imply \eqref{zero-sig}. In addition,
we would expect any 6D Hamiltonian/PS equivalence to be contingent on some condition analogous to the convexity condition required
for the 4D Hamiltonian/Lagrangian equivalence. Following a re-examination of the PS formulation in section \ref{sec:Lagrange}, we propose that the required condition is causality.  

For most of this paper, however,  we rely on the Hamiltonian formulation alone. In  particular, we deduce  all causality conditions on 6D theories of 
chiral 2-form electrodynamics by  examination of the dispersion relations found in  \cite{Bandos:2023yat} for plane-wave perturbations of a general stationary homogeneous background. These dispersion relations were used in  \cite{Bandos:2023yat} to determine the conditions for ``zero trirefringence'' (coincidence of all three dispersion relations).  Here we show how they can be used to re-derive the causality conditions \eqref{causal-h1} and \eqref{causal-h2}, but now in a 6D context.  The method used relies on the simplifying assumption of  a {\sl static} background, which can be justified by the fact that almost all stationary backgrounds are Lorentz boosts of static backgrounds. This method cannot easily accommodate backgrounds 
with $\sigma<0$ but a separate computation, again  using the  dispersion relations of \cite{Bandos:2023yat},  shows that  {\sl any} $\sigma<0$ 
background allows acausal wave propagation. 

This result concludes a proof that the necessary and sufficient conditions for causality of any 6D theory of chiral 2-form electrodynamics
defined by a Hamiltonian CH-function $\hh(\sigma)$ are not only \eqref{causal-h1} and \eqref{causal-h2} but also \eqref{zero-sig}, and this applies equally to 4D self-dual NLED. This is our main result, but  it leaves open the question of the consistency of the $\sigma\ge0$ restriction in those  ($\gamma>0$) cases for which it is relevant. This is because it is in these cases that the physical domain of $\HH$ is restricted, and 
any arbitrarily chosen restriction would almost certainly violate Lorentz invariance.  This issue first arose for the 4D ModMax theory \cite{Bandos:2020jsw}. Here we address the issue more generally with methods applicable to 6D chiral 2-form electrodynamics,  and we argue for consistency based on the impossibility of reaching $\sigma<0$ from $\sigma>0$ by a Lorentz boost.

Using a Lagrangian formulation, we showed in a previous work that the stress-energy tensor of any causal 4D NLED theory satisfies  both the Dominant Energy Condition (DEC) and the Strong Energy Condition (SEC) \cite{Russo:2024xnh}, with much simpler proofs for self-dual NLED. Another aim of this paper is to determine whether these results extend to 6D chiral 2-form electrodynamics, with different methods because we now use a Hamiltonian 
formulation and because the SEC  is a dimension-dependent condition.  The first step is to construct the stress-energy tensor from the Hamiltonian, which we do using methods  developed for 4D NLED in \cite{Bialynicki-Birula:1984daz} and \cite{Mezincescu:2023zny}. Using this result we show that both the DEC and the SEC are indeed a consequence of causality. 

%%%%%%%%%%%%%%%%%%%%%%%%%%%%%%%%%%%%%%%%%%%%%%%%%%
%%%%%%%%%%%%%%%%%%%%%%%%%%%%%%%%%%%%%%%%%%%%%%%%%%
\subsection{4D/6D}\label{subsec1}

We summarize here the dimensional reduction/truncation procedure that converts any given 6D chiral 2-form theory into a 4D self-dual NLED, mostly following \cite{Bandos:2023yat} but with additional comments on  ``consistency'' (in the Kaluza-Klein sense) that are essential for 
the 4D/6D correspondence. 

The dimensional reduction step is accomplished by taking $\bb{A}$ to be independent of $x^2$ and $x^4$. We then set to zero all components of $\bb{A}$ except $\{A_{\alpha 2}, A_{\alpha 4}\, ; \alpha =1,3,5\}$. Without this truncation we would get a 4D self-dual NLED theory coupled to a scalar field. After this procedure we have a Hamiltonian density that is a function of the pair of 3-vector densities
\be 
({\bf B}, {\bf D}) := \{(B^{\alpha 2}, B^{\alpha 4});\,  \alpha=1,3,5\} \, ,  
\ee
and the rotation-invariant 6D variables $(s,p)$ are now the rotation-invariant 4D variables defined by
\be\label{4Dsp}
s= \frac12 \left( |{\bf B}|^2 + |{\bf D}|^2\right) \, , \qquad 
p= |{\bf D} \times {\bf B}|\, .  
\ee
In this 4D interpretation, the variables $(s,p)$ are  invariant under a $U(1)$ electromagnetic duality transformation, equivalent to a phase-shift of the complex 3-vector density ${\bf D}+i{\bf B}$. The 6D Hamiltonian density function $\HH(s,p)$ 
therefore becomes the 4D Hamiltonian density for a self-dual NLED, and we can again use \eqref{uvdefs} to define alternative (but now duality-invariant) variables $(u,v)$. 
Moreover, the ``physical region''  of the $(u,v)$ plane, in which both $u$ and $v$ are now defined in terms of $({\bf D},{\bf B})$, is still $v\ge u\ge 0$.

After the first step (dimensional reduction) the action \eqref{6Dact} is at least quadratic in the fields that are set to zero in the second step (truncation). This is obvious for the free theory and true more generally  because it is true of the scalars $s$ and $p$ as defined in \eqref{6Dsp}. This means that the reduction/truncation is ``consistent'' in the sense that any solution of the 4D NLED field equations is a solution of the 6D field equations with the truncated fields taking their vacuum values. This is significant because it implies that a superluminal wave solution in a constant uniform background of the 4D NLED theory implies a similar solution of the 6D field equations. 

In other words, the reduction/truncation procedure converts any causal 6D chiral 2-form electrodynamics theory to a causal 4D self-dual theory. Since the Hamiltonian formulation used here has both a 4D and 6D interpretation, causality results for 4D self-dual NLED
will also apply to 6D chiral 2-form electrodynamics.

%%%%%%%%%%%%%%%%%%%%%%%%%%%%%%
%%%%%%%%%%%%%%%%%%%%%%%%%%%%%%%%%
%%%%%%%%%%%%%%%%%%%%%%%%%%%%%%%%%
\section{Causality I}\label{sec:waves}
\setcounter{equation}{0}

The full field equations of any (Lorentz invariant) theory of 6D chiral 2-form electrodynamics can be linearized about any solution. A special feature of the class of stationary homogeneous ``background''  solutions is that they preserve the time and space translation invariance of the 6D Minkowski vacuum. This implies that there are plane-wave perturbations with a constant angular velocity $\omega$ and wave 5-vector ${\bf k}$. For the vacuum background the dispersion relation is $\omega^2= |{\bf k}|^2$ for all three polarisation modes. More generally, each of the three dispersion relations takes the form
\be
P_2(\omega, {\bf k}) =0\, , 
\ee
where $P_2$ is a quadratic polynomial in $(\omega, {\bf k})$ with coefficients that are constants but which depend on the choice of polarisation mode. 
For each mode, the polynomial $P_2$ determines $\omega$ as a function of ${\bf k}$ and this yields the group-velocity 
\be
v_g = \left| \frac{\partial\omega}{\partial k}\right|\, , 
\ee
which is now a function of ${\bf k}$ {\sl and} the background (in particular the background momentum density ${\bf p}$. 
A remarkable consequence of Lorentz invariance of the full field equations is that two of the three $P_2$ polynomials coincide \cite{Bandos:2023yat}. There are therefore three polarisations but only two independent dispersion relations. 

If there is a background for which $v_g>1$ for some choice of ${\bf k}$ and polarisation mode then this background `admits' superluminal 
waves, which violates causality\footnote{For some optical materials a superluminal group velocity does not imply a violation of causality, but 
they do not include the backgrounds for any 4D NLED theory (and we expect the same for 6D chiral 2-form electrodynamics); 
 see \cite{Russo:2024kto} for a discussion of this point.}. Any theory with such backgrounds is therefore acausal, unless these backgrounds can be consistently excluded.  Thus, there are two steps in the determination of the class of causal chiral 2-form electrodynamics theories.
The first, is to find the backgrounds that admit superluminal wave propagation, which we address in this and the following section. 
The second,  which we leave to the following section,  is to consider  whether there is a consistent way to exclude some of these backgrounds.

The field equations that follow from \eqref{6Dact} are solved by any constant uniform field $\bb{B}$; these yield the possible stationary homogeneous background solutions. We may orient the axes so that $\bb{B}$ takes the block-diagonal form 
\be\label{blockd}
\bb{B} = B_1 ({\bf e}_1\wedge {\bf e}_2)\,  \oplus \, B_2 ({\bf e}_3\wedge {\bf e}_4)\, , \qquad  B_1\ge B_2\ge0\, . 
\ee
The background values of $s$ and $p$ are then 
\be
s = \frac12(B_1^2+B_2^2)\, , \qquad p= B_1B_2\, , 
\ee
and hence 
\be
B_1^2 = s + \sqrt{s^2-p^2}= 2v \, , \qquad B_2^2 = s- \sqrt{s^2-p^2} =2u\, . 
\ee
For plane-wave perturbations of these backgrounds the residual rotational invariance of $\bb{B}$ allows us to choose axes such that 
\be
k_2=k_4=0\, . 
\ee
The dispersion relations for these plane-waves were found in \cite{Bandos:2023yat} for any Hamiltonian density function $\HH(s,p^2)$. The two independent dispersion relations are
\be\label{pol1}
\!\!\!\!\!\!\! 1.  \qquad \left(\omega + 2k_5 \, p \HH_{p^2}\right)^2 = \HH_s^2 k_5^2 + \HH_s\left( Q_1k_1^2  + Q_2k_3^2\right)\, , 
\ee
and 
\be\label{pol2}
\qquad 2.\qquad  \left(\omega+ k_5 \, p [2\HH_{p^2} + \Lambda]\right)^2 = \Xi_1\Xi_2 k_5^2 +  \Xi_1Q_2 k_1^2 + \Xi_2Q_1 k_3^2 \, . 
\ee
\smallskip

\noindent
Expressions for $\HH_s,\HH_{p^2}$, and the functions $(Q_1,Q_2)$, $(\Xi_1,\Xi_2)$ and $\Lambda$ that were given as functions 
of $(s,p)$ in \cite{Bandos:2023yat} are given below as functions of $(u,v)$: 
\be\label{fds}
\HH_s = \frac{1}{v-u} \left(v\HH_v - u\HH_u\right)\, , \qquad \HH_{p^2}=  \frac{1}{4(v-u)} \left( \HH_u -\HH_v\right)\, . 
\ee
and
\be\label{Qs}
Q_1= \HH_u\, , \qquad Q_2 = \HH_v\, , 
\ee
\be\label{Xis}
\Xi_1 = \HH_u + 2u\HH_{uu}\, , \qquad \Xi_2 = \HH_v + 2v\HH_{vv}\, , 
\ee
and
\be\label{Lambda}
\Lambda = -2\HH_{p^2} + \HH_{uv} = -\frac{(\HH_u-\HH_v)}{2(v-u)}  + \HH_{uv}\, . 
\ee
Other relations, not needed for \eqref{pol1} and \eqref{pol2} but useful elsewhere are
\bea\label{sds}
\HH_{ss}  &=&\  \frac{2uv}{(v-u)^3} \left(\HH_u-\HH_v\right) 
+ \frac{1}{(v-u)^2} \left( v^2\HH_{vv} -2uv \HH_{uv} + u^2\HH_{uu}\right) \nonumber\\
4\HH_{s p^2} &=&\ -\frac{(v+u)}{(v-u)^3} \left(\HH_u-\HH_v\right) 
- \frac{1}{(v-u)^2} \left( v\HH_{vv} -(v+u)\HH_{uv} + u\HH_{uu}\right) \nonumber \\
16\HH_{p^2 p^2} &=&\  \frac{2}{(v-u)^3} \left(\HH_u-\HH_v\right) 
+ \frac{1}{(v-u)^2} \left(\HH_{vv} -2\HH_{uv} + \HH_{uu}\right)\, . 
\eea

All of the above expressions involving partial derivatives of  $\HH(u,v)$  can be converted to expressions 
involving the first and second derivatives of the CH-function $\hh(\sigma)$ by means of \eqref{CH-Ham}. 
The first partial derivatives of $\HH(u,v)$ are given in  \eqref{firstds}.  The second partial derivatives determine
the Hessian matrix for the function $\HH(u,v)$. By using the fact that the definition of $\sigma$ in \eqref{CH-Ham}
implies that\footnote{We use here
the notation of \cite{Russo:2024ptw}.}  
\be\label{tG}
{\tilde G} d\sigma = [\hh^\prime]^3 dv - \hh^\prime du \, , \qquad \tilde G := [\hh^\prime]^3 - 2u h^\prime{}^\prime\, ,  
\ee
one finds that 
\be\label{2nd}
\left(\begin{array}{cc} \HH_{uu} & \HH_{uv} \\ \HH_{vu} & \HH_{vv} \end{array} \right) =  
\frac{\hh^\prime{}^\prime}{\hh^\prime \tilde G} \left(\begin{array}{cc} 1 & -[\hh^\prime]^2 \\ -[\hh^\prime]^2 & [\hh^\prime]^4 \end{array}\right)\, . 
\ee
We shall see later that $\tilde G$ is never zero for a causal theory.

%%%%%%%%%%%%%%%%%%%%%%%%%%%%%%%%%%%%%%%%%%%%%%%%%%%%
%%%%%%%%%%%%%%%%%%%%%%%%%%%%%%%%%%%%%%%%%%%%%%%%%%%%%
\subsection{Static backgrounds}

For almost all backgrounds\footnote{We prove in subsection \ref{subsec:CandC}
that the only exceptions are backgrounds for which $\HH=p$.} it is possible to choose a Lorentz frame such that $p=0$; i.e. a frame in which the background is a {\sl static} homogeneous medium. For this reason, and because it leads to significant simplifications, we will initially restrict to static backgrounds, for which  $u=0$ within the physical region $v\ge u\ge0$.  To proceed we now need  the first and second partial derivatives of 
$\HH(u,v)$ at $u=0$, and hence $\sigma=v$, which can be found from \eqref{firstds} and \eqref{2nd}. This yields
\be\label{uzero1}
\HH_v(0,v) = \hh^\prime(v) \, , \qquad \HH_u (0,v) = 1/\hh^\prime(v)\, , 
\ee
and 
\be\label{uzero2}
\HH_{uu}(0,v) = \frac{\hh^\prime{}^\prime(v)}{[\hh^\prime(v)]^4} \, , \qquad 
\HH_{uv}(0,v) = -  \frac{\hh^\prime{}^\prime(v)}{[\hh^\prime(v)]^2}\, , \qquad 
\HH_{vv}(0,v) = \hh^\prime{}^\prime(v)\, .
\ee
Using these expressions, we find that the dispersion relations of \eqref{pol1} and \eqref{pol2} simplify as follows:
\begin{enumerate}

\item $\omega^2 = A_- (k_5^2 + k_3^2) + k_1^2$, with
\be\label{Aminus}
A_-=  [\hh^\prime(v)]^2 \, . 
\ee

\item $\omega^2 = A_+ (k_5^2 + k_3^2)  + k_1^2$, with
\be\label{Aplus}
A_+ = 1+ \frac{2v\hh^\prime{}^\prime(v)}{\hh^\prime(v)} \, .
\ee
\end{enumerate}

We now have the dispersion relations of any 6D Lorentz invariant theory of chiral  2-form electrodynamics in a form that allows a simple determination of conditions on the function $\hh$ that are 
required for $\HH$ to define a causal theory.

%%%%%%%%%%%%%%%%%%%%%%%%%%%%%%%
%%%%%%%%%%%%%%%%%%%%%%%%%%%%%%%
\subsection{Group velocities}\label{subsec:gv}

To arrive at the dispersion relations of \eqref{Aminus} and \eqref{Aplus} we chose a frame in which 
$p=0$. Since $p=B_1B_2$ and $B_1\ge B_2\ge0$, this frame is one for which $B_2=0$. In other words, 
a static background field $\bb{B}$ has non-zero components only in the ${\bf e}_1\wedge {\bf e}_2$ plane. 
We define ${\bf k}_\parallel$ as the projection of ${\bf k}$ to this plane, 
and ${\bf k}_\perp$ as its orthogonal complement. Recalling that $k_2=k_4=0$ for our choice of axes, we may rewrite the dispersion relations 
more generally as
\be\label{static-disp}
\omega^2= A_\pm |{\bf k}_\perp|^2 + |{\bf k}_\parallel |^2\, .
\ee
The corresponding group velocities are 
\be\label{groupv}
{\rm v}_g^\pm = \sqrt{ \frac{A_\pm^2 |{\bf k}_\perp|^2 +  |{\bf k}_\parallel |^2}
{A_\pm |{\bf k}_\perp|^2 + |{\bf k}_\parallel |^2}}\, ,  
\ee
where the $\pm$ superscript labels the two independent dispersion relations. 
In order for these group velocities to be defined and non-superluminal for all ${\bf k}$, we require
\be\label{Apm}
0 \le A_\pm \le 1\,. 
\ee

This is essentially the result found previously for 4D NLED theories in \cite{Bialynicki-Birula:1984daz,Russo:2022qvz}  but there $A_\pm$ was found in terms of the 4D Lagrangian density and its derivatives (with respect to Lorentz scalars). 
In the current Hamiltonian context (4D and 6D) we have $A_\pm$ expressed in terms of the CH-function $\hh$ according to the equations \eqref{Aminus} and \eqref{Aplus}. Using these equations, and taking into account that $\hh^\prime >0$, we see that \eqref{Apm} is satisfied for $A_-$ iff\footnote{Causality permits $A_-=0$, which is implied by $\hh^\prime=0$, but \eqref{CH-Ham} requires $\hh^\prime \ne0$.}
\be\label{c1}
0 <\hh^\prime \le 1\, . 
\ee
Similarly, we see (since $\hh^\prime>0$ and $v\ge0$) that $A_+\le1$ requires
\be\label{c2}
\hh^\prime{}^\prime \le 0\, .
\ee
Notice that \eqref{c1} and \eqref{c2} imply that $\tilde G>0$, as claimed earlier. This leaves the $A_+>0$ condition,  
which becomes the following inequality
\be\label{c3}
\hh^\prime + 2v \mathfrak{h}^\prime{}^\prime \ge 0\, . 
\ee
Finally, noting that the argument of the function $\hh$ appearing in \eqref{c1}, \eqref{c2} and \eqref{c3} is the {\sl non-negative} variable $v$, 
we see that we may replace $v$ by $\sigma$ to arrive at the inequalities of  \eqref{causal-h1} and \eqref{causal-h2}  in the Introduction provided that we also include the restriction \eqref{zero-sig} on $\sigma$; i.e. $\sigma\ge0$. 

To summarise: by examination of the dispersion relations for plane-wave perturbations of {\sl static} homogeneous background solutions of the general chiral 2-form electrodynamics theory, with Hamiltonian density $\HH(u,v)$ expressed in terms of the CH-function $\hh(\sigma)$, we have recovered
the causality conditions on $\hh$ given in the Introduction for 4D self-dual NLED; i.e.  \eqref{causal-h1} and \eqref{causal-h2}, {\sl and}  the restriction  \eqref{zero-sig}. This is expected from  the 4D/6D correspondence but the derivation here, directly from 6D dispersion relations, is much simpler. 

However, we have not yet learnt anything about backgrounds with $\sigma<0$. 
This is the  topic of the following section.

%%%%%%%%%%%%%%%%%%%%%%%%%%%%%%%%%
%%%%%%%%%%%%%%%%%%%%%%%%%%%%%%%%%
%%%%%%%%%%%%%%%%%%%%%%%%%%%%%%%%
\section{Causality II}\label{sec:C2}
\setcounter{equation}{0}

In order to understand what distinguishes  $\sigma<0$ backgrounds (when they exist) from $\sigma\ge0$ backgrounds,
we will first need to understand what is special about $\sigma=0$. Some methods of analysis used for this purpose in the 4D context 
\cite{Russo:2024ptw} can be taken over without change to the 6D context. For example, from the definition of $\sigma$ in \eqref{CH-Ham} we see  that each value of $\sigma$ is associated with a particular straight line in the $(u,v)$ plane, with slope  
\be\label{slope}
{\rm Slope}(\sigma) =  \left[\hh^\prime(\sigma)\right]^{-2}  \, . 
%= \frac{(v-\sigma)}{u}\, .
\ee
We have also seen that causality requires both $\hh'(\sigma)\le1$ and $\hh''(\sigma)\le 0$ for 
$\sigma\ge0$. These conditions ensure that the lines of constant $\sigma\ge0$ foliate
a region in the positive quadrant of the $(u,v)$-plane, but the nature of this region 
depends crucially on whether $\gamma=0$ or $\gamma>0$ (we ignore $\gamma<0$ because we 
already know that this leads to acausal theories). We focus first on $\gamma=0$, leaving $\gamma>0$ for the following subsection. 

%%%%%%%%%%%%%%%%%%%%%%%%%%%%%%%%%%%%%%%%%%%%
\begin{figure}[h!]
 \centering
\includegraphics[width=0.5\textwidth]{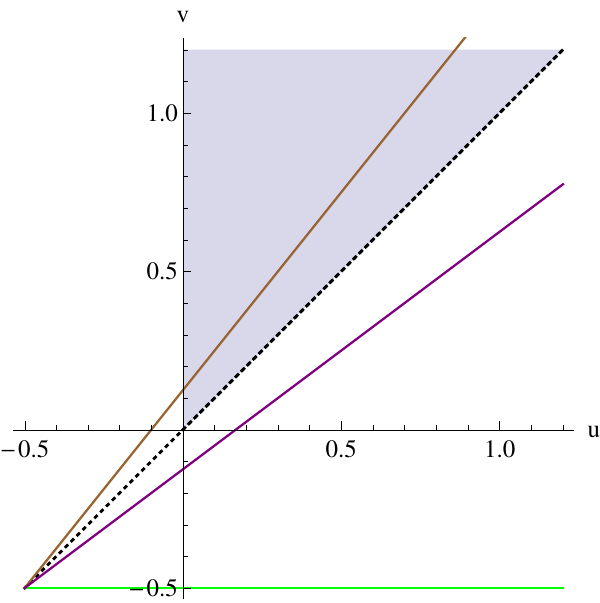}
  \caption{Lines of constant $\sigma$ for the `BI/M5' theory. The shaded region is the ``physical region'' $v\geq u\geq 0$ for which $(u,v)$ are defined in terms of the fields. The line in the shaded area is a line of constant $\sigma>0$. The horizontal green line corresponds to the (negative) minimum value of $\sigma$ for which $\hh(\sigma)$ is real. All $\sigma<0$ lines (such as the purple line) lie entirely outside the physical region.}
 \label{fig1}
\end{figure}
%%%%%%%%%%%%%%%%%%%%%%%%%%%%%%%%%%%%%%%%%%%%%%

The simplest $\gamma=0$ case is the free-field theory with $\hh(\sigma)=\sigma$. In this case the region foliated by lines of constant $\sigma\ge0$ is the entire physical region 
because the $\sigma=0$ line is (for $u\ge0$) its $v=u$ boundary line. This will remain true for any causal theory with the same weak-field limit except that the lines of constant $\sigma$ will no longer be parallel,  which leads to intersections outside the physical region. This is illustrated in Fig. \ref{fig1} for the BI/M5 case. A special feature of this case is that all the lines intersect at the same point in the negative quadrant (generically they intersect on a caustic rather than a focus).  A more important feature is that 
{\sl all lines of constant $\sigma<0$ lie entirely outside the physical region}. This is because ${\rm Slope}(\sigma) \le 1$ for $\sigma<0$, which is true because 
\be\label{negsig}
\hh'(\sigma)\ge1 \, \quad (\sigma<0). 
\ee
We should expect this to be true generically, at least in a neighbourhood of 
$\sigma=0$, because causality requires the inequality $\hh'{}'(0) \le 0$, which  is
not generically saturated. 

Whenever \eqref{negsig} {\sl is} true then $\sigma<0$ is not physically realisable (for $\gamma=0$) and the $\sigma\ge0$ condition does not restrict the (physically-realisable) domain of $\HH(u,v)$.  If \eqref{negsig} is {\sl not} true then some lines of constant  $\sigma<0$ will pass through the physical region and hence intersect some lines of constant $\sigma>0$ in this region. At these intersections the function $\HH(u,v)$ ceases to be uniquely defined, and this will restrict the domain of $\HH(u,v)$. This is not a violation of causality {\it per se} but it could lead to a violation of Lorentz invariance. 

Another feature of the BI/M5 case illustrated in Fig. \ref{fig1} is that 
${\rm Slope}(\sigma)$ is a non-decreasing function of $\sigma$ because 
$\hh'{}'\le0$ not only for $\sigma\ge0$ (as required for causality) but also
for $\sigma<0$. For simplicity, we shall assume here the following ``extended concavity'' 
condition
\be
\forall \sigma :\ \hh'{}'(\sigma) \le 0 \, ,  
\ee
which implies \eqref{negsig} for all causal theories.  

The main point here is that {\sl there are no backgrounds with $\sigma<0$ when $\gamma=0$}, 
so we turn to $\gamma>0$.

%%%%%%%%%%%%%%%%%%%%%%%%%%%%%%%%%%%%%%%%%%%%%%%%%%%%%%%%%
%%%%%%%%%%%%%%%%%%%%%%%%%%%%%%%%%%%%%%%%%%%%%%%%%%%%%%%%%%
\subsection{Why $\gamma>0$ requires $\sigma\ge0$}\label{subsec:g<0}

For $\gamma>0$ the $\sigma=0$ line is no longer the lower boundary of the ``physical region''. Instead, it is the line $v= e^{2\gamma}u$, which separates the physical region into two disjoint subregions:
\begin{itemize}
         \item Upper subregion: $R_+ = \{u\ge 0, \, v>e^{2\gamma}u\}$.

         \item Lower subregion: $R_- = \{ u>0, \, e^{2\gamma}u>v \ge u\}$.
\end{itemize}
%%%%%%%%%%%%%%%%%%%%%%%%%%%%%%%%%%%%%%%%%%%%%%%%%%%%%%%%%%%
\begin{figure}[h!]
 \centering
\includegraphics[width=0.5\textwidth]{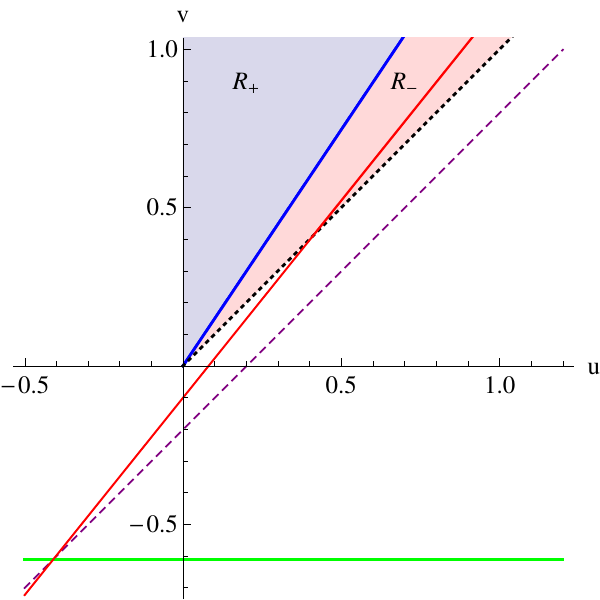}
  \caption{Constant $\sigma$ lines in $(u,v)$ space for ModMaxBorn and its 6D analog.
  Above the dotted line ($v=u$) the fields are real, but all lines of constant $\sigma>0$ now lie in the blue shaded $R_+$ region. The red line is one with $\sigma<0$ that intersects 
  $v=u$. The purple dashed line is parallel to the $v=u$ line. 
 }
 \label{fig22}
\end{figure}
%%%%%%%%%%%%%%%%%%%%%%%%%%%%%%%%%%%%%%%%%%%%%%%%%%%%%%%%%%
Let us first consider the simplest case (ModMax or its 6D analog):
\be\label{sigzero}
\hh(\sigma) = e^{-\gamma}\sigma\, \qquad (\gamma>0). 
\ee
The subregion $R_+$ is foliated by half-lines of $\sigma>0$ whereas $R_-$ is foliated by half-lines of $\sigma<0$. Allowing all values of $\sigma$ we get the function 
\be\label{MM}
\HH(u,v) = e^{-\gamma} v + e^\gamma u \qquad (\gamma>0), 
\ee
defined in the entire physical region ($v\ge u\ge0$). In this particular case $\sigma<0$ {\sl is} physically realisable (in $R_-$) and this will be true for any theory with \eqref{MM} as its conformal weak-field limit. This is illustrated in fig. \ref{fig22} for the ``ModMaxBorn''  4D NLED \cite{Bandos:2020jsw} 
 and its 6D analog \cite{Bandos:2020hgy} for which 
\be\label{hMMB}
h(\sigma) = \sqrt{T^2 + 2Te^{-\gamma} \sigma} -T\, , \qquad 
(\gamma>0).
%\qquad \left(\Rightarrow \quad \hh^\prime = \frac{e^{-\gamma}}{\sqrt{1+2e^{-\gamma}\sigma/T}}\right)\,    
\ee

Our analysis of wave propagation in static backgrounds does not apply to backgrounds
associated to points in $R_-$ because $p\ne0$ in $R_-$. They can be viewed as boosts of static backgrounds with $v=0$ (rather than $u=0$) on the ``second branch''  of the physical region that we discussed briefly in the Introduction. However, it is simpler to return to the dispersion relations of 
\eqref{pol1} and \eqref{pol2} for generic stationary homogeneous backgrounds. It will suffice to consider \eqref{pol1}, which we shall do for two choices of the wave-vector ${\bf k}$: orthogonal and parallel to the background momentum vector-density ${\bf p}$:

For orthogonal wave propagation (${\bf k}\cdot{\bf p}=0$) we set $k_5=0$ in \eqref{pol1}. 
Using both \eqref{fds} and \eqref{Qs}, we then find that\footnote{It should be appreciated   ``orthogonal''  is being used here in a different sense to subsection \ref{subsec:gv}.  To recover the result found there for ``orthogonal'' wave propagation in a static background  one must set $u=0$ (which implies that $\sigma=v$) and $k_3=0$, to get $\omega^2= k_1^2$.} 
\be\label{perp}
\omega^2 = \left(\frac{\sigma}{v-u}\right) \left(k_1^2 + \left[\hh'(\sigma)\right]^2 k_3^3\right)\, . 
\ee
We see immediately that $\omega^2<0$ for $\sigma<0$, which indicates an instability of all backgrounds corresponding to points in $R_-$. For $\sigma=0$ we have $\omega=0$, i.e. no instability but also no wave propagation. 

For parallel wave propagation ($|{\bf k}\times {\bf p}|=0$) we set $k_1=k_3=0$ in \eqref{pol1}. This yields
\be
\omega = \left(-2p\HH_{p^2} \mp \HH_s\right) k_5 = 
\mp \left(\frac{\sqrt{v}\HH_v \pm \sqrt{u}\HH_u}{\sqrt{v} \pm \sqrt{u}}\right)k_5\, , 
\ee
where \eqref{fds} has been used for the second equality. The group velocities are therefore 
\be
v_g^\pm = \left| \frac{[\hh'(\sigma)]^2\pm \sqrt{u/v}}{ \hh^\prime(\sigma)\left(1 \pm \sqrt{u/v}\right)}\right|\, . 
\ee
Let us focus on $v_g^+$. The condition $v_g^+\le 1$ is equivalent to
\be\label{c-interval}
\hh'(\sigma) \in \left[\sqrt{u/v}, 1\right]\, . 
\ee
We already know, for $\sigma\ge0$, that causality requires $\hh'(\sigma)\le 1$. We now learn that this causality condition applies more generally. More importantly for the issue at hand, we learn that causality requires $\hh'\ge \sqrt{u/v}$. By comparing this inequality with the relation \eqref{hp2} (which defines $\sigma$) we see that causality requires $\sigma\ge 0$. 

This shows that for theories with $\gamma>0$ causality requires us to impose the restriction $\sigma\ge0$ on the domain of $\hh(\sigma)$, which implies a restriction of the domain of $\HH(u,v)$ to $R_+$. But is this restriction consistent?

%%%%%%%%%%%%%%%%%%%%%%%%%%%%%%%%%%%%%%%%%%%%%%
%%%%%%%%%%%%%%%%%%%%%%%%%%%%%%%%%%%%%%%%%%%%%%%
\subsection{Convexity and Consistency}\label{subsec:CandC}

The issue of consistency of the restriction of the domain of $\HH(u,v)$ to the $R_+$ subregion of the physical region in the $(u,v)$ plane is not restricted to 6D chiral 2-form electrodynamics. It applies equally to 4D self-dual NLED and it  is useful to first consider its significance in this context where $\HH(u,v)$ can be written as a function of the 3-vector densities $({\bf D}, {\bf B})$.  A simple requirement of causality in this case is that $\HH$ should be a convex function of ${\bf D}$. This requires positivity (no negative eigenvalues) of the Hessian matrix  
\be
({\rm Hess})_{ij} := \frac{\partial^2 \HH}{ \partial D_i \partial D_j}\, .
\ee
When $\HH$ is expressed as a function of the variables $(s,p^2)$, with their 4D interpretation, a calculation yields
\bea\label{Hess}
({\rm Hess})_{ij} &=& \left(\HH_s + 2|{\bf B}|^2\HH_{p^2}\right)\delta_{ij} -2\HH_{p^2} B_iB_j \\
&&+\ \HH_{ss} D_iD_j -4\HH_{sp^2} D_{(i} ({\bf p}\times {\bf B})_{j)} 
+4\HH_{p^2p^2} ({\bf p}\times {\bf B})_i ({\bf p}\times {\bf B})_j\, . \nonumber
\eea 
Results of \cite{Mezincescu:2023zny}, where the eigenvalues were computed on the weaker assumption that $\HH$ is rotation invariant but not necessarily duality invariant, imply
(but we omit the details here) that a necessary condition for positivity of the Hessian matrix is that $\HH_s\ge0$ (as was found for ModMax in \cite{Bandos:2020jsw}). 
From  \eqref{fds}, we see that this inequality is equivalent to 
\be
0 \le v\HH_v - u\HH_u = \left[\hh^\prime\right]^{-1} \left(\left[\hh^\prime\right]^2v -u\right)\, , 
\ee
which is equivalent to the following condition on $\hh^\prime$:
\be\label{hp1}
\hh^\prime(\sigma) \ge \sqrt{\frac{u}{v}}\, . 
\ee
By comparison with the definition of $\sigma$ in \eqref{CH-Ham} rewritten as
\be\label{hp2}
\hh^\prime(\sigma) = \sqrt{\frac{u}{v-\sigma}}\, , 
\ee
we see that the convexity condition $\HH_s\ge0$ is equivalent to $\sigma\ge0$.
Thus, for 4D self-dual NLED the subregion $R_+$ in the $(u,v)$ plane is the 
``convex domain'' of $\HH$ (the domain for which it is a convex function of ${\bf D}$). 
This is true for $\gamma\ge0$. In this 4D context the domain of $\HH$ is necessarily its convex domain if it is defined via a Legendre transform (as verified explicitly for ModMax in \cite{Bandos:2020jsw}) and this is a compelling argument for the consistency of the restriction to $R_+$.  However, it is not applicable 
(in any obvious way) to 6D chiral 2-form electrodynamics. 

We now present a further argument for consistency that applies equally for both 4D and 6D contexts. It is based on the observation that consistency with Lorentz invariance requires that no point in $R_-$ can be reached by a Lorentz boost of any point in $R_+$. Arbitrarily chosen restrictions on the domain of $\HH$ in the $(u,v)$ plane are almost certainly inconsistent with Lorentz invariance because almost all (constant uniform) backgrounds are boosts of some static background. In fact, a result of\cite{Mezincescu:2023zny} is that the only stationary backgrounds that are {\sl not} Lorentz boosts of a static background are those for which\footnote{This formula is equivalent, for $\HH(u,v)$, to $W=p$ of \cite{Mezincescu:2023zny}.} 
\be
v\HH_v + u\HH_u =p \, . 
\ee
Squaring both sides we find that this is equivalent to $(v\HH_v- u\HH_v)^2=0$, which is equivalent to $H_s=0$, and hence to $\sigma=0$. Any boost of any point on the $\sigma=0$ line results in another point on this line, which is therefore Lorentz invariant. 

It follows that no boost of a point in $R_+$ can ever reach $R_-$ because to do so it would have to cross the $\sigma=0$ line. At best, the $\sigma=0$ line could be approached asymptotically by an infinite succession of boosts. Starting from any point in $R_+$, a sufficiently large Lorentz boost will change the values of both $\HH$ and $p$ such that 
$\HH\approx p$, which suggests that $\HH=p$ at $\sigma=0$. This is true and it can be proved by using \eqref{hp2} (and the fact that $p^2=4uv$) to rewrite \eqref{CH-Ham} as 
\be
\HH(u,v) = \hh(\sigma) + \sqrt{p^2 -4u\sigma}\, ,  \qquad 
\sigma= v- \frac{u}{[\hh'(\sigma)]^2} \, . 
\ee
If we assume that $\hh(0)=0$ (zero vacuum energy), and that $\hh(\sigma) \to e^{-\gamma}\sigma$ in the weak-field limit, then we have
\be\label{zerosig}
\HH\big|_{\sigma=0} = p\, , \qquad v= e^{-2\gamma} u\, . 
\ee

It was shown in \cite{Russo:2022qvz,Mezincescu:2023zny} that small-amplitude plane-wave perturbations of a background with $\HH=p$ can propagate only in the direction of the background flow, which is consistent with the observation of the previous subsection that there is no wave propagation 
in a direction orthogonal to ${\bf p}$ in a  background with $\sigma=0$.

To summarise: we have now shown that the restriction on the physical domain of $\HH(u,v)$ that is required for causality when $\gamma>0$ is 
consistent with Lorentz invariance. This result applies equally to 4D self-dual NLED.

%%%%%%%%%%%%%%%%%%%%%%%%%%%%%%%
%%%%%%%%%%%%%%%%%%%%%%%%%%%%%%%
%%%%%%%%%%%%%%%%%%%%%%%%%%%%%%%
\section{Stress-energy tensor}
\setcounter{equation}{0}

The field equation that follows from the variation of $\bb{A}$ in the 
action of \eqref{6Dact} is
\be\label{eom}
\dot{\bb{B}} = \boldsymbol{\nabla} \times \bb{H}\, ,  
\ee
where $\bb{H}$ was defined in \eqref{defH}.
This field equation implies the continuity equations 
\be
\partial_\mu \TT^{\mu 0} =0  \, , \quad \partial_\mu \TT^{\mu j}=0 \qquad (\mu=0,i)\, , 
\ee
where the components of the $\TT$-tensor are defined as 
\be
\TT^{00} = \HH\, , \qquad \TT^{i0} = - \left(\bb{H}\times\bb{H}\right)^i\, , 
\qquad \TT^{0j} = -\left(\bb{B}\times\bb{B}\right)^i\, , 
\ee
and
\be
\TT^{ij} = B^{ki} H^j{}_k + \delta^{ij} \left(2W - \HH\right)\, , 
\ee
where
\be\label{defH}
W=  \frac12 \bb{B}\cdot \bb{H}\, , \qquad \bb{H} := \frac{\partial\HH}{\partial \bb{B}}\, .
\ee
The derivative with respect to  $\bb{B}$  is defined\footnote{Definitions in the literature can differ by a factor of 2.} such that $\bb{H}=\bb{B}$ for the free-field theory with $\HH= \frac12 |\bb{B}|^2$.  This 5-space stress-tensor is not obviously symmetric. However, rotational invariance implies that 
$\HH$ can be expressed as a function of $(s,p^2)$, in which case 
\be\label{bbH}
\bb{H} = \HH_s \bb{B} - 2\HH_{p^2} \left({\bf p}\times \bb{B}\right)\, . 
\ee
The identities
\be
  B^k{}_{[i} B_{j]k} \equiv0\, , \qquad  B^k{}_{[i} ({\bf p}\times \bb{B})_{j]k} \equiv 0\, , 
\ee
may then be used to deduce that  $B^{k[i} H^{j]}{}_k=0$, and hence that $\TT^{ij}= \TT^{ji}$.

Observe that $\TT^{i0}$ is not necessarily the same as $\TT^{0i}$, but it {\sl is} the same for 
a Lorentz invariant theory because the condition for Lorentz invariance is \cite{Bandos:2020hgy}
\be
\bb{H}\times\bb{H} = \bb{B}\times\bb{B} \, . 
\ee
In this case we can identify $\TT^{\mu\nu}$ as the symmetric stress-energy tensor of the chiral 2-form electrodynamics 
theory defined by $\HH$. 

At any chosen point in the spacetime we may choose axes such that $\bb{B}$ takes the  form of \eqref{blockd}. It then follows from \eqref{bbH} that $\bb{H}$ takes a similar block-diagonal form, with 
\be
\begin{aligned}
H_1 := H_{12} &=\  \HH_s B_1 + 2p\HH_{p^2} B_2= \sqrt{2v}\HH_v \\
H_2 := H_{34} &=\ \HH_s B_2 + 2p\HH_{p^2} B_1 = \sqrt{2u}\HH_u\, . 
\end{aligned}
\ee
This implies that $H_1H_2= p$ as a consequence of Lorentz invariance (as expected because $B_1B_2=p$). It also implies that 
\be
B_1 H_1 = 2v \HH_v \, , \qquad  B_2 H_2 = 2u\HH_u\, , 
\ee
and hence that 
\be\label{Wagain}
W = v\HH_v + u\HH_u \, .
\ee
Using the Lorentz invariance condition \eqref{HamLI}, we find that 
\be
W^2 -p^2 = (u\HH_u-v\HH_v)^2 \ge 0\, . 
\ee
If we also define
\be\label{Wpm} 
W_+ := 2v\HH_v\, , \qquad W_- := 2u\HH_u\, , 
\ee
then we can write $W$ as 
\be
W= \frac12 (W_+ +W_-) \, , 
\ee
with 
\be
W_\pm = W \pm \sqrt{W^2-p^2}\, . 
\ee
Finally, when $\HH$ is expressed in terms of $\hh$ according to \eqref{CH-Ham} then both $\HH_u$
and $\HH_v$ are positive, which implies that $W_\pm$ are both non-negative, and 
\be
W\ge p\, . 
\ee
Notice that
\be\label{Wineq}
2W \ge W_+ \ge W_- \ge0\, . 
\ee

From these results and definitions, we see that the stress-energy tensor  takes the following block-diagonal form
after the rows and columns are permuted to take the order $051234$:
\be\label{set6D}
\TT = \left(\begin{array}{cccc} \HH & p & 0 & 0\\ p& 2W - \HH & 0 &0\\
0 & 0 & [W_- - \HH] \bb{I}_2 & 0\\ 0& 0 & 0 & [W_+ -\HH] \bb{I}_2 \end{array} \right)\, .
\ee
This is essentially the result for generic 4D NLED found and discussed in  \cite{Mezincescu:2023zny}, except
that the two diagonal $2\times2$ blocks (the 12 and 34 planes) become two $1\times1$ blocks in the 4D case, 
as one would expect from dimensional reduction in the 2 and 4 directions. The expressions of \eqref{Wagain} 
and \eqref{Wpm} for $W$ and $W_\pm$ are also applicable in 4D but only to self-dual NLED.

Taking the trace of the stress-energy tensor with the 6D Minkowski metric we find that 
\be
\Theta := \TT^{\mu\nu} \eta_{\mu\nu} = 6(W-\HH)\, .
\ee
This is zero when $\HH(u,v)$ is a degree-1 homogeneous function of  $(u,v)$, which is therefore
the condition for conformal invariance.  Another scalar that we can construct from the stress-energy tensor is 
\be
\label{tttt}
\TT^2 := \TT^{\mu\nu}\TT_{\mu\nu} = 6\left[(W - \HH)^2 + (W^2-p^2)\right] \, . 
\ee
These expressions for $\Theta $ and $\TT^2$ are equivalent to those found previously in \cite{Ferko:2024zth}. 
Notice that the expression for $\TT^2$   in \eqref{tttt}  makes clear 
the fact that
the {\sl unique} chiral-2-form theory for which both Lorentz scalars $\Theta$ and $\TT^2$ are zero is the one with $W=\HH =p$. 
As explained earlier, this is the one case for which $\HH$ cannot be expressed in terms of the one-variable function $\hh$.

We  now turn to an investigation of the implications of causality on the energy conditions satisfied by the 
stress-energy tensor of \eqref{set6D}. We shall consider in turn the Dominant Energy Condition (DEC) and 
the Strong Energy Condition (SEC).

%%%%%%%%%%%%%%%%%%%%%%%%%%%%%%%%%%%
%%%%%%%%%%%%%%%%%%%%%%%%%%%%%%%%%%%%
\subsection{The DEC}\label{subsec:DEC}

The Dominant Energy Condition (DEC) is equivalent to the statement that $\HH$ must be greater
than or equal to the absolute value of all components of the stress-energy tensor. We therefore require, in particular, 
that $\HH \ge |2W-\HH|$,  which implies (since $W$ cannot be negative) that $\HH\ge W$. This implies that
$\HH\ge p$, because $W\ge p$. The inequalities \eqref{Wineq} ensure that $\HH$ similarly  ``dominates'' all other  terms. 
The DEC is therefore equivalent to 
\be\label{concavity}
\HH\ge W  \qquad (DEC).
\ee
If we choose $\HH(u,v)$ to be zero in the vacuum, i.e. 
\be\label{vacE}
\HH(0,0) = 0 \, , 
\ee
then \eqref{concavity} is equivalent to the statement that $\HH(u,v)$ is a concave function\footnote{ This applies  equally for the 4D interpretation, which is remarkable when one considers that,  in this context,  causality also requires $\HH$ to be a {\sl convex} function of ${\bf D}$.}, which 
is the case iff the Hessian matrix of $\HH(u,v)$ is negative (non-positive eigenvalues). This matrix is 
given in of \eqref{2nd}; its two eigenvalues $\lambda$ are
\be\label{evals}
\lambda = \left\{ 0 \, , \ \left(\frac{1+ [\hh^\prime]^4}{\hh^\prime\tilde G}\right)\hh^\prime{}^\prime \right\} \, . 
\ee
The DEC is therefore satisfied  iff $\hh^\prime{}^\prime\le0$, which is precisely the causality condition of \eqref{c2}. 
Thus, causality implies the DEC.

%%%%%%%%%%%%%%%%%%%%%%%%%%%%%%%%%%%%%%%%%%%%%
%%%%%%%%%%%%%%%%%%%%%%%%%%%%%%%%%%%%%%%%%%%
\subsection{The SEC}

For a $D$-dimensional Minkowski spacetime, with Minkowski metric $\eta$,  the SEC requires, for any timelike vector field $\xi$, that
\be
\xi^\mu \xi^\nu \left[ \mathcal{T}_{\mu\nu} - \frac{1}{(D-2)} \Theta \eta_{\mu\nu} \right] \ge 0\, .  
\ee
At any point in spacetime we can choose coordinates $x^\mu= (t,x^i)$ such that $\xi\propto \partial_t$, in which case the SEC is
\be
\HH + \frac{1}{(D-2)} \Theta\geq 0\, . 
\ee
For the $D=4,\, 6$ cases of interest here, we know that $\Theta= D(W-\HH)$ and hence the SEC is equivalent to
\be 
DW - 2\HH \ge0\, , 
\ee
which we can rewrite as 
\be\label{SEC}
\tfrac12 (D-4)W + (2W-\HH) \ge 0 \qquad (SEC).
\ee

Let us first consider a 4D self-dual NLED, with a Hamiltonian density function $\HH(u,v)$. The 
SEC is then the condition
\be
0 \le 2W-\HH=2v\HH_v + 2u\HH_u -\HH =: K(u,v)\, . 
\ee 
Using \eqref{CH-Ham} and \eqref{firstds}, we find that
\be
K(u,v) = 2v\, \hh^\prime(\sigma) - \hh(\sigma) \, , \qquad \sigma = v - \frac{u}{[\hh^\prime(\sigma)]^2}\, .
\ee
A calculation shows that 
\be
\begin{aligned}
K_u(u,v) &=\,  \frac{1}{\tilde G} \left[ [\hh^\prime]^2 - 2v\hh^\prime{}^\prime\right]\, , \\
K_v(u,v) &=\,  \frac{\hh^\prime}{\tilde G} \left[ [\hh^\prime]^2 (\hh^\prime + 2v\hh^\prime{}^\prime) - 4u \hh^\prime \right]\, . 
\end{aligned}
\ee
Both of these partial derivatives are non-negative for all $(u,v)$ as a consequence of the causality conditions \eqref{c1}, \eqref{c2} {\sl and} \eqref{c3}, 
so the lowest possible value of $K$ for any causal theory is 
\be
K(0,0) = - \hh(0) =0\, . 
\ee
The SEC is therefore satisfied, and saturated in the vacuum. 

\medskip

Next, we consider the 6D case. Now the SEC condition \eqref{SEC} is 
\be
W+K\ge0\ ,
\ee
but since $W\ge0$ and $K\ge0$ (and both $W$ and $K$ are zero in the vacuum)
we conclude that the SEC inequality is again satisfied and saturated in the vacuum.  

%%%%%%%%%%%%%%%%%%%%%%%%%%%%%%%%%%%%%%%%
%%%%%%%%%%%%%%%%%%%%%%%%%%%%%%%%%%%%%%%%%
%%%%%%%%%%%%%%%%%%%%%%%%%%%%%%%%%%%%%%%%
\section{A Lagrangian perspective}\label{sec:Lagrange}
\setcounter{equation}{0}

In previous sections we have made occasional use of the correspondence between 4D theories of self-dual NLED and 
6D  theories of  chiral 2-form electrodynamics, which  is a consequence of the fact that any given Hamiltonian density 
function $\HH(u,v)$ has both a 4D and a 6D interpretation.  Is there a `Lagrangian' version of this 4D/6D correspondence? 

There is a possible correspondence of the 4D Lagrangian formulation of self-dual NLED with the Perry-Schwarz (PS) formulation 
of 6D chiral 2-form electrodynamics \cite{Perry:1996mk}, which can be viewed as a  `pseudo-Hamiltonian' formulation with manifest 
invariance under the 5D Lorentz group $SO(1,4)$ rather than the 6D rotation group $SO(5)$. The analog of the 
phase-space action is the integral of the ``PS-Lagrangian'' density:
\be\label{PS}
\LL_{\rm PS} = \frac14 A^\prime_{mn} B^{mn} - \HH_{\rm PS}(B)\, , 
\ee
where $A$ is now a 2-form potential on 5D Minkowski spacetime with field-strength dual
\be
B^{mn} = \frac12 \varepsilon^{mnpqr} \partial_p A_{qr}\, , 
\ee
but it is also a function of a fifth space coordinate,  and $A^\prime$ indicates the derivative of $A$ with respect to it. The `pseudo-Hamiltonian'  
density $\HH_{\rm PS}$ is a 5D Lorentz scalar function of $B$, which may be written as  function of the following two 5D Lorentz scalars \cite{Perry:1996mk}:
\be
y_1 = \frac12 {\rm tr} B^2  , \qquad y_2= \frac14 {\rm tr} B^4 \, , 
\ee
where the matrix $B$ has entries $B_m{}^n = \eta_{mp}B^{pn}$. 

The fact that $4y_2\ge y_1^2$ suggests a different basis for 5D Lorentz scalars that was also considered in \cite{Perry:1996mk}:
\be\label{PSchoice}
u_\pm = - \frac12 \left(y_1 \pm \sqrt{4y_2 -y_1^2} \right) \, . 
\ee
For our purposes the following slightly different version  is preferable\footnote{The factor of $\frac12$ is needed because $\HH_{PS}$ is one half of the 
corresponding function used in \cite{Perry:1996mk}.}:
\be\label{6Dint}
(U,V) = \frac12(u_-,-u_+)\, . 
\ee
For this choice of basis the condition found in \cite{Perry:1996mk} 
for the field equations of \eqref{PS} to be 6D Lorentz invariant is
\be\label{GR}
\LL_U\LL_V =-1\, , \qquad \LL(U,V) := \HH_{\rm PS}(U,V)\, . 
\ee

If we now dimensionally reduce by setting $A'=0$ then $\LL_{\rm PS}$ becomes the Lagrangian 
density $\LL_{\rm 5D}:=  -\HH_{\rm PS}$ for a generic 5D electrodynamics theory, and this becomes the Lagrangian 
density for a 4D self-dual NLED theory upon further dimensional reduction followed by a truncation. Details of this 
reduction/truncation to 4D can be found in \cite{Bandos:2020hgy}. The 6D variables $(U,V)$ become 
\be\label{4Dint}
U= \frac12\left(\sqrt{S^2+P^2} -S\right) \, , \qquad  V= \frac12\left(\sqrt{S^2+P^2} + S\right) \, , 
\ee
where $(S,P)$ are the standard scalar and pseudo-scalar 4D Lorentz invariants 
constructed from the electric and magnetic components of the 2-form field-strength:
\be
S= \frac12 (|{\bf E}|^2 - |{\bf B}|^2)\, , \qquad P= {\bf E}\cdot {\bf B}\, . 
\ee
In this 4D context, $\LL(U,V)$ becomes the (manifestly Lorentz invariant) Lagrangian density for a 4D NLED, with \eqref{GR} the condition for $U(1)$ electromagnetic duality invariance of its field equations \cite{Gibbons:1995cv}; i.e. self-duality. 

Thus, $\LL(U,V)$ defines either a  4D self-dual NLED theory or a 6D chiral electrodynamics theory  according to the interpretation given 
to the variables $(U,V)$. This is similar to the status of the Hamiltonian density $\HH(u,v)$, but there is a previously unappreciated difference.
It is obvious from \eqref{4Dint} that the variables $(U,V)$ are both non-negative for their 4D interpretation, and this implies that the  4D ``physical region''  in the $(U,V)$ plane is its positive quadrant\footnote{This assumes that the square-root terms in \eqref{4Dint} are positive rather than negative; for the opposite sign choice the physical region would be the negative quadrant.}. Using the 6D definitions of $(U,V)$ in terms of $(y_1,y_2)$, we find that 
\be
S = \frac12 y_1 \, , \qquad P^2 = y_2 - \frac12 y_1^2\, . 
\ee 
We noted above that $4y_2 -y_1^2\ge0$ but since this inequality can be saturated (a proof by example is given in the following subsection)
it follows that $2y_2-y_1^2$ may be positive, zero or negative. If it is negative then there are 6D field configurations such that $P^2<0$ for the 6D 
interpretation of $P$. As this is not possible for the 4D interpretation, the range of  the variables $(S,P^2)$ is larger for 6D, and the same is therefore true of the variables $(U,V)$. We turn now to a precise determination of this  6D ``physical region''. 

%%%%%%%%%%%%%%%%%%%%%%%%%%%%%%%%%%%%%%%%%
%%%%%%%%%%%%%%%%%%%%%%%%%%%%%%%%%%%%%%%%%
\subsection{The PS physical region}\label{subsec:PS+}

To fully determine the possible values of $(U,V)$ in  their 6D interpretation, it is convenient to use the 4-space rotational invariance of $(U,V)$ to bring the $5\times5$ matrix $B$ to the following form\footnote{This is not a ``canonical'' form because  we use only the rotation part of the manifest $SO(1,4)$ symmetry group, but it suffices for our purposes.} for which the non-zero entries in the first row and column are electric-field components:
\be
B= \left(\begin{array}{ccccc} 0 & E_1 & 0 & E_2 & 0 \\ E_1 & 0 & B_1 & 0 & 0 \\ 0 & -B_1 & 0 &0 &0 \\ 
E_2 & 0 & 0 & 0 & B_2\\ 0&0&0& -B_2 & 0\end{array} \right)\, \qquad (B_1\ge B_2). 
\ee
For this matrix $B$,  a computation yields
\be
y_1 = X+Y\, , \qquad 4y_2-y_1^2 = (X+Y)^2  + 4(E_1^2E_2^2 - XY)\, , 
\ee
where
\be
X= E_1^2-B_1^2 \, , \qquad Y = E_2^2-B_2^2\, . 
\ee
In this notation, we see that 
\be 
4U = \sqrt{(X+Y)^2  + 4(E_1^2E_2^2 - XY)} - (X+Y)\, .
\ee
This is obviously non-negative when $X+Y \le 0$. It is still non-negative when $X+Y >0$ because of the 
identity
\be 
\left(E_1^2E_2^2 -XY\right) \left(B_1^2+B_2^2\right)  \equiv  (X+Y)(B_1B_2)^2 + \left(E_1^2B_2^4 + E_2^2 B_1^4\right)\, ,  
\ee
which implies that $E_1^2E_2^2 \ge XY$ when $X+Y>0$.
A similar argument does {\sl not} apply to 
\be
4V= \sqrt{(X+Y)^2  + 4(E_1^2E_2^2 - XY)} + (X+Y)\, , 
\ee
which is obviously non-negative when $X+Y\ge0$, but may be positive or negative when $X+Y<0$. In fact, 
\be
4(V+U) = \sqrt{ (X-Y)^2 + 4(E_1E_2)^2} \ \ge 0\, , 
\ee
with equality when $X=Y$ and $E_1E_2=0$. For example\footnote{This example confirms the claim made earlier that the inequality 
$4y_2-y_1^2\ge0$ can be saturated.}, if  we choose $E_2=0$ then $V=-U$ when $E_1^2= B_1^2-B_2^2$ (in which case $X+Y =-2B_2^2$).

The physical region for the 6D variables $(U,V)$ is therefore the intersection of the regions $U\ge0$ and $V+U\ge 0$. This is larger than the 4D physical region because it includes points with $V<0$. As for the Hamiltonian formulation, there  is a distinction to be made between the two boundary half-lines of the physical region; in this case $U=0$ for $V\ge0$ and $U+V=0$ for $U\ge0$. The latter half-line is where  the two branches of $(U,V)$ meet, as mentioned earlier. These two branches are exchanged by the involution
%\footnote{We called this ``$\Phi$-parity'' in \cite{Russo:2024ptw} but that was in the context of 4D self-dual NLED.}
\be\label{PhiP}
(U,V) \to -(V,U)\, , 
\ee
which has $V+U=0$ as a fixed line\footnote{It was argued in \cite{Perry:1996mk} that this involution should be a symmetry 
of $\LL(U,V)$ because this would ensure a free-field theory 
in the weak-field limit (i.e. $\gamma=0$) but it is a stronger condition than $\gamma=0$.}. 

The implications of this result for the Lagrangian 4D/6D correspondence are not immediately clear. 
We address this issue in the following subsection.

%%%%%%%%%%%%%%%%%%%%%%%%%%%%%%%
%%%%%%%%%%%%%%%%%%%%%%%%%%%%%%
\subsection{CH-construction for PS}

For either the 4D or  6D interpretation of $\LL(U,V)$, we have the following solution of \eqref{GR}
in terms of a Lagrangian CH-function $\ell(\tau)$ \cite{Perry:1996mk,Russo:2024ptw}:
\be\label{CH-Lag}
\LL(U,V) = \ell(\tau) - \frac{2U}{\dot\ell(\tau)} \, , \qquad  \tau= V + \frac{U}{[\dot\ell(\tau)]^2}\,  \qquad (\dot\ell>0),  
\ee
where the overdot indicates a derivative with respect to $\tau$. These equations imply that 
\be\label{Lfds}
\LL_V = \dot\ell\, , \qquad \LL_U = -1/\dot\ell \, , 
\ee
and hence that \eqref{GR} is solved. From  its definition, we see that each value of $\tau$ corresponds to a straight line 
in the $(U,V)$ plane with slope 
\be
{\rm Slope}(\tau) = - \left[\dot\ell(\tau)\right]^{-2}\, ,  
\ee
which is the Lagrangian analog of \eqref{slope}.  

In order to have a weak-field expansion we assume that $\ell(\tau)$ has a Taylor expansion of the form 
\be\label{Taylor'}
\ell(\tau) = e^\gamma \tau + \mathcal{O}(\tau^2)\, , 
\ee
where $\gamma$ is a constant. By comparing the weak-field limit for the 4D Lagrangian and Hamiltonian formulations, this constant can be identified as the constant $\gamma$ appearing in the Taylor expansion of \eqref{Taylor} for $\hh(\sigma)$.  The linear term by itself yields the general conformal weak-field limit, and there are again significant differences between the $\gamma=0$ and $\gamma>0$ cases. 

%%%%%%%%%%%%%%%%%%%%%%%%%%%%%%%%%%%%%%%%%%%%%%%%%%%
\begin{figure}[h!]
 \centering
 \begin{tabular}{cc}
\includegraphics[width=0.4\textwidth]{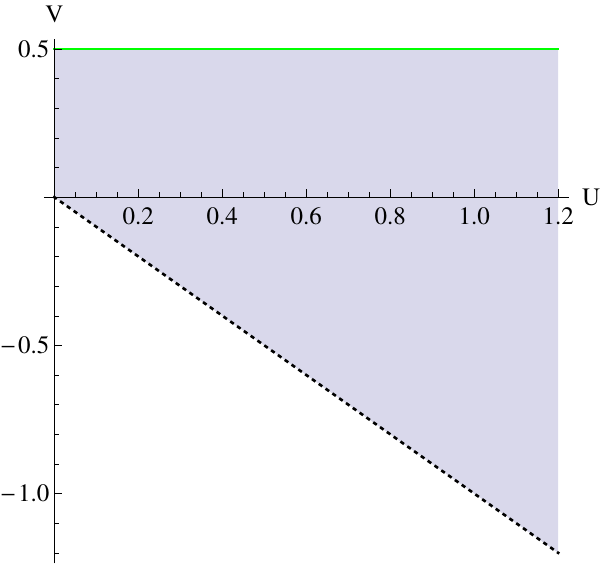}
 &
 \qquad \includegraphics[width=0.4\textwidth]{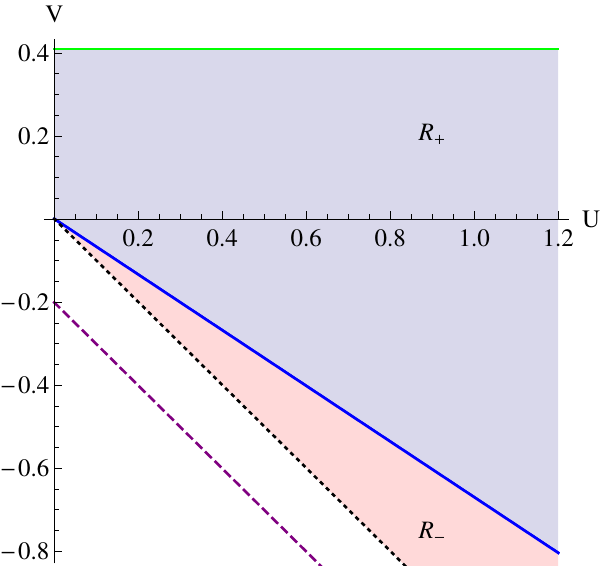}
 \\ (a)&(b)
 \end{tabular}
 \caption{a) The `M5' theory ($\gamma=0$) the physical region is foliated by constant $\tau$ half-lines $0<\tau<T/2$. The green line is $2\tau=T$. 
 b) The 6d analog of ModMaxBorn ($\gamma>0$). The blue line $V+ e^{-2\gamma}U=0$ separates the subregions $R_+$ ($\tau>0$) and $R_-$ ($\tau<0$) of the physical region.  
 The purple (dashed) line ($\tau=-T\sinh\gamma$) is parallel to the $V+U=0$ boundary of $R_-$.
 }
 \label{fig:tau}
 \end{figure}
%%%%%%%%%%%%%%%%%%%%%%%%%%%%%%%%%%%%%%%%%%%%%%%%%%%

For simplicity, we focus on the conformal weak-field limits and comment briefly on the effects of interactions.  For the free-field ($\gamma=0$) case we have $\LL(U,V)= V-U$ and the entire physical region in the $(U,V)$ plane is foliated by either segments of lines of constant $\tau\ge0$ (in the 4D case) or half-lines of $\tau\ge0$ (in the 6D case).  Although, these 4D and 6D regions differ, the CH-function $\ell(\tau)$ is {\sl exactly the same, not only in its values but also in its domain}. Interactions do not change this. This is illustrated in fig. \ref{fig:tau}a for the `M5' theory (6D analog of BI) found from the CH-function 
\be\label{ellM5}
\ell(\tau)= T- \sqrt{T^2 - 2T\tau}\, , 
\ee
There is now a maximum value for $\tau$ and this implies a maximum value of $V$. The domain of
$\LL(U,V)$ in the 6D case is again larger than the 4D domain, but the domain of $\ell(\tau)$ is the same: $0<2\tau<T$. 
This implies a one-to-one 4D/6D Lagrangian correspondence, at least for  the  ($\gamma=0$) subclass of theories that 
become free-field theories in the weak-field limit. 

For the $\gamma>0$ conformal weak-field limit we have
\be
\LL(U,V) = e^{-\gamma}V - e^\gamma U\, ,  
\ee
and the $\tau=0$ line no longer coincides with the $V+U=0$ boundary of the 
6D physical region. Instead, it splits this region into two disjoint subregions 
\begin{itemize}
         \item Upper subregion: $R_+ = \{U\ge 0, \, V >-e^{2\gamma}U\}$.

         \item Lower subregion: $R_- = \{U>0, \, -e^{2\gamma}U>V \ge -U\}$.
\end{itemize}
The entire physical region is foliated by lines of constant $\tau$ but $R_+$ is foliated by $\tau\ge0$ half-lines while $R_-$ is foliated by 
$\tau<0$ half-lines.  Thus, the domain of $\ell(\tau)$ for 6D is now {\sl larger} than it is for 4D {\sl unless we restrict the domain of $\LL(U,V)$ to $R_+$}, since this is  equivalent to $\tau\ge0$. With this restriction we again have a 4D/6D correspondence, again based on a single function $\ell(\tau)$, with a definite domain, that has both a 4D and 6D interpretation. 

This conclusion is not changed by interactions. For example, the CH-function $\ell(\tau)$ for the ModMaxBorn self-dual NLED discussed in section \ref{sec:C2} is 
\be
\ell(\tau) = T- \sqrt{T^2 - 2e^{\gamma} T\tau}\, , 
\ee
which applies equally to the  6D analog of ModMaxBorn,  illustrated  in Fig.\ref{fig:tau}b. The (dark-shaded) subregion $R_+$ is now foliated by lines of constant $\tau\ge0$ up to a maximum value, and the (light-shaded) subregion $R_-$ is foliated by the lines of constant $\tau<0$, up to a minimum value of $\tau$ specified by the bound ${\rm Slope}(\tau)<-1$.  Again the domain of $\ell(\tau)$ is now larger for 6D than it was for 4D, unless we 
restrict the domain of $\LL(U,V)$ to $R_+$. 

To summarise: For $\gamma=0$ there is a Lagrangian 4D/6D correspondence despite different domains of $\LL(U,V)$ in its 4D and 6D interpretations, but for this to remain true for $\gamma>0$ we must restrict the domain of $\ell(\tau)$ to $\tau\ge0$. This restriction has no effect for 4D, or for 6D when $\gamma=0$, 
but it restricts the function $\LL(U,V)$ to subregion $R_+$ of the 6D physical region when $\gamma>0$.

%%%%%%%%%%%%%%%%%%%%%%%%%%%%%%%%%%%%%
%%%%%%%%%%%%%%%%%%%%%%%%%%%%%%%%%%%%
\subsection{Causality for PS} 

We explained in subsection \ref{subsec1} why causality conditions on 4D self-dual NLED are necessary conditions for causality of 
6D chiral 2-form electrodynamics because the former is a consistent reduction/truncation of the latter. That was in the context of a Hamiltonian formulation but we have just seen how it can be extended to the Lagrangian/PS formulations. We therefore begin with a brief review of causality conditions for 4D self-dual NLED. They are very simple if we assume that $\LL(U,V)$ has a weak-field expansion about 
a conformal weak-field limit: it is both necessary and sufficient\footnote{For generic NLED one additional ``strong-field'' causality condition is needed \cite{Schellstede:2016zue}; see also \cite{Russo:2024kto}.} for $\LL$ to be a strictly convex function of the electric field ${\bf E}$.  This convexity condition on $\LL$ as a function of ${\bf E}$ can be traded for conditions on the function $\LL(S,P)$ \cite{Bandos:2021rqy};  one of them is $\LL_S>0$, which is required for other reasons detailed in \cite{Schellstede:2016zue}. The remaining conditions are  equivalent to convexity of the function $\LL(S,P)$.

When $\LL$ is rewritten in terms of the 4D Lorentz scalar variables $(U,V)$, the condition $\LL_S>0$ becomes equivalent 
(provided $V+U\ne0$) to $V\LL_V-U\LL_U>0$. In terms of the CH-function $\ell(\tau)$ this is 
\be\label{ccc1} 
V\dot\ell + U/\dot\ell >0 \, 
\ee
which is satisfied\footnote{For 4D. We revisit this condition below for 6D.} since $\dot\ell>0$ and both $U$ and $V$ are non-negative. Thus, causality is equivalent to convexity of $\LL(S,P)$, which was  
shown in \cite{Russo:2024llm} to be equivalent to the following simple causality conditions on the CH-function $\ell(\tau)$:
\be\label{causality}
\dot\ell \ge1 \, , \qquad \ddot\ell\ge 0\, . 
\ee
Since $\HH$ may be found from $\LL$ by a Legendre transform, the CH-functions $\hh(\sigma)$ and $\ell(\tau)$ must be related, and they are \cite{Russo:2024ptw}. The conditions on $\ell(\tau)$ are simpler because $\tau$ is necessarily non-negative for 4D self-dual NLED. 

The condition $\ddot\ell\ge0$, which states that $\ell(\tau)$ is a convex function, is equivalent to the statement that $\LL(U,V)$ is convex\footnote{Convexity of $\LL(U,V)$ does not imply convexity of $\LL(S,P)$ because the latter condition implies {\sl both} equations of \eqref{causality}.}. This follows from results of \cite{Russo:2024llm} that can be written in matrix form as\footnote{This is the Lagrangian analog of \eqref{Hess}.}
\be\label{HessL}
\left(\begin{array}{cc} \LL_{UU} & \LL_{UV} \\ \LL_{VU} & \LL_{VV} \end{array} \right) =  
\frac{\ddot\ell} {\dot \ell  G} \left(\begin{array}{cc} 1 & \dot \ell^2 \\ \dot \ell^2 & \dot \ell^4 \end{array}\right)\, ,\qquad
G :=\dot\ell^3+2U\ddot \ell\ .  
\ee
The two eigenvalues $\lambda$ are 
\be\label{evalsL}
\lambda = \left\{ 0 \, , \ \left(\frac{1+ \dot \ell^4}{\dot \ell G}\right)\ddot \ell\right\} \, ,  
\ee
which are both non-negative iff $\ddot\ell\ge0$. 

We now turn to the 6D interpretation of $\LL(U,V)$ within the PS formulation of chiral 2-form electrodynamics. We first reconsider the significance of \eqref{ccc1}; taking into account that $V<0$ is now possible, we may rewrite it for $V<0$ as 
\be\label{ccc2}
\dot\ell < \sqrt{\frac{U}{|V|}}\, , \qquad (0 >V > -U),  
\ee
Comparing this with the definition of $\tau$ in \eqref{CH-Lag}, rewritten for $V<0$ as
\be
\dot\ell = \sqrt{\frac{U}{\tau+ |V|}}\, ,
\ee
we see that \eqref{ccc1} (which is  $\LL_S>0$)  is now equivalent to $\tau>0$ (and that $\LL_S=0$ on the 
line $\tau=0$). This is analogous to the result established in subsection \ref{subsec:CandC} that $\HH_s\ge0$ is equivalent to $\sigma\ge0$, with $\HH_s=0$ on the $\sigma=0$ line; as in that case we must distinguish 
between the two  possible conformal weak-field limits:
\begin{itemize}
    \item $\ell(\tau) = \tau + \mathcal{O}(\tau^2) \qquad (\gamma=0)$

    \item $\ell(\tau) = e^\gamma\tau + \mathcal{O}(\tau^2) \qquad (\gamma>0)$
\end{itemize}

For $\gamma=0$ the $\tau=0$ line is the boundary half-line $V=-U$ (for $U>0$) of the physical region, which therefore coincides with the region $R_+$ foliated by lines of constant $\tau>0$.  Furthermore, under conditions analogous to those spelled out in our discussion of causality in the Hamiltonian formulation (but which we pass over here) {\sl all  lines with $\tau<0$ lie entirely outside the physical region}. This allows us to `lift' to 6D the causality conditions of \eqref{causality}.

The analysis for $\gamma>0$ is more involved because $\tau<0$ is physically realisable; it corresponds to the region $R_-$.  We expect it to be excluded by causality;  this should not be difficult to check but it would require a re-derivation of dispersion relation is the PS formalism, which is outside the scope of this paper. Irrespective of the result, $R_-$ must be excluded in order for there to be Lagrangian 4D/6D equivalence, and then the causality conditions \eqref{causality} will apply in 6D. The more important issue is therefore consistency of the restriction to $R_+$. It is not clear to us how it can be addressed within the PS formulation, but we expect results that are in accord with our discussion of this issue for the Hamiltonian formulation.

%%%%%%%%%%%%%%%%%%%%%%%%%%%%%%%%%%%%%
%%%%%%%%%%%%%%%%%%%%%%%%%%%%%%%%%%%%
%%%%%%%%%%%%%%%%%%%%%%%%%%%%%%%%%%%%%%%%
\section{Summary and Outlook}
\setcounter{equation}{0}

We have explored various features of the general Lorentz invariant  theory of chiral two-form electrodynamics in a 6D Minkowski spacetime. Our main aim has been to determine the conditions required for causal propagation of plane-wave perturbations of generic stationary homogeneous backgrounds, and we have addressed this issue mainly using the Hamiltonian formulation. This has the advantage that it is very close to the Hamiltonian formulation of 4D self-dual NLED, to which it reduces via a process of dimensional reduction/truncation. 

For both 4D and 6D the general rotation invariant Hamiltonian density $\HH$ can be chosen to be a function of two independent rotation-invariants $(u,v)$ defined in terms of the fields (4D or 6D) for the same ``physical region'' in the $(u,v)$-plane. Any choice of $\HH(u,v)$ satisfying $\HH_u\HH_v=1$ defines a Lorentz invariant theory for both 4D self-dual NLED and 6D chiral 2-form electrodynamics, implying 
a 4D/6D correspondence  \cite{Bandos:2020hgy}. An apparent obstacle to this correspondence is that linearisation of the Hamiltonian field equations about an arbitrary stationary and homogeneous background solution yields a wave equation with plane wave solutions that have three independent polarisations for 6D but only two for 4D. This is resolved by the fact that only two of the three dispersion relations in 6D can differ, and these become the two dispersion relations of the corresponding 4D theory \cite{Bandos:2023yat}. 

This condition for $\HH(u,v)$ to define a Lorentz invariant theory can be solved in terms of a one-variable ``CH-function'' 
$\hh(\sigma)$, which (omitting one limiting case) parameterises the possible interacting Lorentz-invariant theories with the same degrees of freedom as the free theory \cite{Russo:2024ptw}, and this applies for both the 4D and 6D interpretations. The 4D/6D equivalence is now reduced to the statement that a single function $\hh$ determines both a 4D and an a 6D theory.  Starting from the expressions found in \cite{Bandos:2023yat}  for the two independent 6D dispersion relations, we have  found causality conditions on $\hh$ by requiring causal propagation in static backgrounds. They coincide with conditions that we previously deduced for self-dual NLED  \cite{Russo:2024ptw} by 
`translating'  Lagrangian results of \cite{Russo:2024llm}. 

As is customary, we have assumed the existence of a weak-field expansion of the Hamiltonian density, with a conformal weak-field limit. This assumption is equivalent to supposing that $\hh(\sigma)$ has a Taylor series expansion of the form 
\be
\hh(\sigma) = e^{-\gamma} \sigma + \mathcal{O}(\sigma^2)\, . 
\ee
The linear term determines the conformal weak-field limit, which is acausal for $\gamma<0$ \cite{Bandos:2020jsw,Bandos:2020hgy}, so we assume $\gamma\ge0$. For $\gamma=0$ the weak-field limit is a free-field theory and for $\gamma>0$ it is ModMax, for 4D, or its 6D chiral 2-form analog. Although the function $\hh(\sigma)$ is defined for $\sigma<0$ (at least near $\sigma=0$) its values for $\sigma<0$ are not physically realisable, and our causality conditions on $\hh$ are both necessary and sufficient. 

This is much less clear for $\gamma>0$ because there are physical Hamiltonian field configurations with $\sigma<0$. Static uniform background fields are naturally divided unto those with $\sigma>0$ and those with $\sigma<0$, with the causality conditions on $\hh(\sigma)$ applying only for $\sigma\ge0$. For the 4D interpretation this inequality is equivalent to a convexity condition on $\HH$ that is known to be required for causality. There is no obvious analog of this argument for 6D chiral 2-form electrodynamics but it is not needed for $\gamma=0$ because
there are no physical backgrounds with $\sigma<0$ in these cases. For $\gamma>0$ there are physical backgrounds with $\sigma<0$, but 
we have shown, using the dispersion relations of \cite{Bandos:2023yat},  that all such  backgrounds allow superluminal propagation.

This result completes the necessary and sufficient conditions for causality for 6D chiral 2-form electrodynamics except for an issue of consistency of the $\sigma\ge0$ condition in the $\gamma>0$ case. Arguments for the consistency of this restriction were made for ModMax 
in \cite{Bandos:2020jsw}. We have presented a new consistency argument here, applicable for both 4D and 6D,  based on the observation that field configurations with $\sigma>0$ cannot be obtained from those with $\sigma<0$ by any Lorentz boost. 

In an earlier work we showed that any causal NLED (with zero vacuum energy) has a stress-energy tensor that satisfies both the Dominant Energy Condition (DEC) and (more surprisingly) the Strong Energy Condition (SEC) \cite{Russo:2024xnh}. This is true for self-dual NLED in particular, and the proofs are much simpler in this case, at least in the Lagrangian formulation. Using the Hamiltonian formulation, 
and different methods, we have extended this result to 6D chiral 2-form electrodynamics. 

Finally, we have revisited the Perry-Schwarz (PS) formulation of 6D chiral 2-form electrodynamics \cite{Perry:1996mk} in which only 5D Lorentz invariance is manifest and interactions are introduced via a PS-function of 5D Lorentz invariants. There is a dimensional-reduction/truncation procedure that takes this function to the Lagrangian density of a 4D self-dual NLED theory, and a single function $\LL(U,V)$ that defines
both the 4D and the 6D theory. This suggests a Lagrangian version of the 4D/6D correspondence, which is supported by the fact that
$\LL(U,V)$ can be expressed in terms of a Lagrangian CH-function $\ell(\tau)$.  However, a previously unnoticed fact is that the physical
domain of $\LL(U,V)$ is larger for  6D, as are the domains of $\ell(\tau)$ for $\gamma>0$. We have argued that a $\tau\ge0$ condition, similar to the Hamiltonian $\sigma\ge0$, must be imposed in this case. If it is imposed then the 4D and 6D domains of $\ell(\tau)$ agree and the
Lagrangian 4D/6D correspondence is confirmed. 

This result has implications for the proposed 6D Hamiltonian/PS correspondence mentioned in the introduction, because it provides an alternative route. If we start with the Hamiltonian density for a particular theory of 6D chiral 2-form electrodynamics we get a 4D self-dual 
NLED Hamiltonian density $\HH$ by the (Hamiltonian) 4D/6D correspondence. Provided that $\HH$ has appropriate convexity properties, 
a Legendre transform gives us the  4D Lagrangian density $\LL$ describing the same self-dual NLED theory. In the final step, we can use the Lagrangian 4D/6D correspondence to get a PS description of the initial 6D theory. Indirectly, this would imply a Hamiltonian/PS equivalence
that is contingent on the convexity properties needed for the involutive property of the Legendre transform at the 4D level.  What is the 6D description of this contingency?  Here, the fact that the convexity properties are equivalent to causality for 4D self-dual NLED provides the answer: it must be causality. We are thus led to the conclusion that there is a 6D Hamiltonian/PS correspondence but it applies only to causal theories. 

We conclude with a few comments on possible applications of our results to other areas of theoretical physics. Causal 6D theories of 2-form electrodynamics are candidates for consistent bosonic truncations of $(1,0)$ supersymmetric theories; we expect $(2,0)$ supersymmetry to allow only the `M5' theory but that $(1,0)$ supersymmetry will be less restrictive. If so, it is likely that they will be relevant to 5-branes in string/M-theory. A related topic is compactifications of $(1,0)$ supersymmetric extensions of chiral 2-form electrodynamics. Is there a one-to-one correspondence with $\mathcal{N}=2$ supersymmetric self-dual NLED theories? Compactifications from 6D to 2D also deserve to be explored. Finally, we remark that the $U(1)$ duality invariance group of 4D self-dual NLED will be broken to a $Z_2$ duality group in the quantum theory, and that this $Z_2$ symmetry is an example of Born-reciprocity \cite{Born:1938zve}, which has more recently been argued to be an important feature of string theory \cite{Freidel:2013zga}.

%%%%%%%%%%%%%%%%%%%%%%%%%%%%%%%%%%%%%%%%%
%%%%%%%%%%%%%%%%%%%%%%%%%%%%%%%%%%%%%%%%%
\section*{Acknowledgements}

JGR acknowledges financial support from grant 2021-SGR-249 (Generalitat de Catalunya) and  by the Spanish  MCIN/AEI/10.13039/501100011033 grant PID2022-126224NB-C21.

%\appendix

%%%%%%%%%%%%%%%%%%%%%%%%%%%%%% References %%%%%%%%%%%%%%%%%%%%%%

\providecommand{\href}[2]{#2}\begingroup\raggedright\endgroup

%\bibliographystyle{fullsort}
%\bibliography{etmgbib}

\begin{thebibliography}{10}

%\cite{Howe:1983fr}
\bibitem{Howe:1983fr}
P.~S.~Howe, G.~Sierra and P.~K.~Townsend,
``Supersymmetry in Six-Dimensions,''
Nucl. Phys. B \textbf{221} (1983), 331-348
%doi:10.1016/0550-3213(83)90582-5

%\cite{Howe:1996yn}
\bibitem{Howe:1996yn}
P.~S.~Howe and E.~Sezgin,
``D = 11, p = 5,''
Phys. Lett. B \textbf{394} (1997), 62-66
%doi:10.1016/S0370-2693(96)01672-3
[arXiv:hep-th/9611008 [hep-th]].


%\cite{Perry:1996mk}
\bibitem{Perry:1996mk}
M.~Perry and J.~H.~Schwarz,
``Interacting chiral gauge fields in six-dimensions and Born-Infeld theory,''
Nucl. Phys. B \textbf{489} (1997), 47-64
%doi:10.1016/S0550-3213(97)00040-0
[arXiv:hep-th/9611065 [hep-th]]. 

%\cite{Pasti:1997gx}
\bibitem{Pasti:1997gx}
P.~Pasti, D.~P.~Sorokin and M.~Tonin,
``Covariant action for a D = 11 five-brane with the chiral field,''
Phys. Lett. B \textbf{398} (1997), 41-46
%doi:10.1016/S0370-2693(97)00188-3
[arXiv:hep-th/9701037 [hep-th]].

%\cite{Howe:1997fb}
\bibitem{Howe:1997fb}
P.~S.~Howe, E.~Sezgin, P.~C.~West and M.~Dine,
``Covariant field equations of the M-theory five-brane,''
Phys. Lett. B \textbf{399} (1997), 49-59
%doi:10.1201/9781482268737-20
[arXiv:hep-th/9702008 [hep-th]].

%\cite{Howe:1997vn}
\bibitem{Howe:1997vn}
P.~S.~Howe, E.~Sezgin and P.~C.~West,
``The Six-dimensional selfdual tensor,''
Phys. Lett. B \textbf{400} (1997), 255-259
%doi:10.1016/S0370-2693(97)00365-1
[arXiv:hep-th/9702111 [hep-th]].


%\cite{Bandos:1997ui}
\bibitem{Bandos:1997ui}
I.~A.~Bandos, K.~Lechner, A.~Nurmagambetov, P.~Pasti, D.~P.~Sorokin and M.~Tonin,
``Covariant action for the superfive-brane of M theory,''
Phys. Rev. Lett. \textbf{78} (1997), 4332-4334
%doi:10.1103/PhysRevLett.78.4332
[arXiv:hep-th/9701149 [hep-th]].

%\cite{Siegel:1983es}
\bibitem{Siegel:1983es}
W.~Siegel,
``Manifest Lorentz Invariance Sometimes Requires Nonlinearity,''
Nucl. Phys. B \textbf{238} (1984), 307-316
%doi:10.1016/0550-3213(84)90453-X

%\cite{Pasti:1996vs}
\bibitem{Pasti:1996vs}
P.~Pasti, D.~P.~Sorokin and M.~Tonin,
``On Lorentz invariant actions for chiral p forms,''
Phys. Rev. D \textbf{55} (1997), 6292-6298
%doi:10.1103/PhysRevD.55.6292
[arXiv:hep-th/9611100 [hep-th]].


%\cite{Avetisyan:2022zza}
\bibitem{Avetisyan:2022zza}
Z.~Avetisyan, O.~Evnin and K.~Mkrtchyan,
``Nonlinear (chiral) p-form electrodynamics,''
JHEP \textbf{08} (2022), 112
%doi:10.1007/JHEP08(2022)112
[arXiv:2205.02522 [hep-th]].

%\cite{Henneaux:1988gg}
\bibitem{Henneaux:1988gg}
M.~Henneaux and C.~Teitelboim,
``Dynamics of Chiral (Selfdual) $P$ Forms,''
Phys. Lett. B \textbf{206}, 650-654 (1988)
%doi:10.1016/0370-2693(88)90712-5

%\cite{Bandos:2020hgy}
\bibitem{Bandos:2020hgy}
I.~Bandos, K.~Lechner, D.~Sorokin and P.~K.~Townsend,
``On p-form gauge theories and their conformal limits,''
JHEP \textbf{03} (2021), 022
%doi:10.1007/JHEP03(2021)022
[arXiv:2012.09286 [hep-th]].


%\cite{Bialynicki-Birula:1984daz}
\bibitem{Bialynicki-Birula:1984daz}
I.~Bialynicki-Birula,
``Nonlinear Electrodynamics: Variations on a theme by Born and Infeld'', 
in {\sl Quantum Theory of Particles and Fields}, eds. B. Jancewicz and 
J. Lukierski, (World Scientific, 1983) pp. 31-48. 


%\cite{Bandos:2020jsw}
\bibitem{Bandos:2020jsw}
I.~Bandos, K.~Lechner, D.~Sorokin and P.~K.~Townsend,
``A non-linear duality-invariant conformal extension of Maxwell's equations,''
Phys. Rev. D \textbf{102} (2020), 121703
%doi:10.1103/PhysRevD.102.121703
[arXiv:2007.09092 [hep-th]].



%\cite{C&H}
\bibitem{C&H}
R. Courant and D. Hilbert, ``Methods of Mathematical Physics'', Vol.II  
(Wiley Interscience, 1962) pp.91-94. 

%\cite{Russo:2024ptw}
\bibitem{Russo:2024ptw}
J.~G.~Russo and P.~K.~Townsend,
``Dualities of self-dual nonlinear electrodynamics,''
JHEP \textbf{09}, 107 (2024)
%doi:10.1007/JHEP09(2024)107
[arXiv:2407.02577 [hep-th]].


%\cite{Bialynicki-Birula:1992rcm}
\bibitem{Bialynicki-Birula:1992rcm}
I.~Bialynicki-Birula,
``Field theory of photon dust,''
Acta Phys. Polon. B \textbf{23}, 553-559 (1992)

%\cite{Gibbons:2000ck}
\bibitem{Gibbons:2000ck}
G.~W.~Gibbons and P.~C.~West,
``The Metric and strong coupling limit of the M5-brane,''
J. Math. Phys. \textbf{42}, 3188-3208 (2001)
%doi:10.1063/1.1376158
[arXiv:hep-th/0011149 [hep-th]].

%\cite{Townsend:2019ils}
\bibitem{Townsend:2019ils}
P.~K.~Townsend,
``An interacting conformal chiral 2-form electrodynamics in six dimensions,''
Proc. Roy. Soc. Lond. A \textbf{476}, no.2236, 20190863 (2020)
%doi:10.1098/rspa.2019.0863
[arXiv:1911.01161 [hep-th]].

%\cite{Russo:2024llm}
\bibitem{Russo:2024llm}
J.~G.~Russo and P.~K.~Townsend,
``Causal self-dual electrodynamics,''
Phys. Rev. D \textbf{109}, no.10, 105023 (2024)
%doi:10.1103/PhysRevD.109.105023
[arXiv:2401.06707 [hep-th]].

%\cite{Ferko:2024zth}
\bibitem{Ferko:2024zth}
C.~Ferko, S.~M.~Kuzenko, K.~Lechner, D.~P.~Sorokin and G.~Tartaglino-Mazzucchelli,
``Interacting chiral form field theories and $ T\overline{T} $-like flows in six and higher dimensions,''
JHEP \textbf{05} (2024), 320
%doi:10.1007/JHEP05(2024)320
[arXiv:2402.06947 [hep-th]].


%\cite{Bandos:2023yat}
\bibitem{Bandos:2023yat}
I.~Bandos, K.~Lechner, D.~Sorokin and P.~K.~Townsend,
``Trirefringence and the M5-brane,''
JHEP \textbf{06}, 171 (2023)
%doi:10.1007/JHEP06(2023)171
[arXiv:2303.11485 [hep-th]].



%\cite{Russo:2024xnh}
\bibitem{Russo:2024xnh}
J.~G.~Russo and P.~K.~Townsend,
``Causality and energy conditions in nonlinear electrodynamics,''
JHEP \textbf{06}, 191 (2024)
%doi:10.1007/JHEP06(2024)191
[arXiv:2404.09994 [hep-th]].


%\cite{Mezincescu:2023zny}
\bibitem{Mezincescu:2023zny}
L.~Mezincescu, J.~G.~Russo and P.~K.~Townsend,
``Hamiltonian birefringence and Born-Infeld limits,''
JHEP \textbf{02}, 186 (2024)
%doi:10.1007/JHEP02(2024)186
[arXiv:2311.04278 [hep-th]].

%\cite{Russo:2024kto}
\bibitem{Russo:2024kto}
J.~G.~Russo and P.~K.~Townsend,
``Born again,''
SciPost Phys. \textbf{16}, no.5, 124 (2024)
%doi:10.21468/SciPostPhys.16.5.124
[arXiv:2401.04167 [hep-th]].



%\cite{Russo:2022qvz}
\bibitem{Russo:2022qvz}
J.~G.~Russo and P.~K.~Townsend,
``Nonlinear electrodynamics without birefringence,''
JHEP \textbf{01} (2023), 039
%doi:10.1007/JHEP01(2023)039
[arXiv:2211.10689 [hep-th]].



%\cite{Gibbons:1995cv}
\bibitem{Gibbons:1995cv}
G.~W.~Gibbons and D.~A.~Rasheed,
``Electric - magnetic duality rotations in nonlinear electrodynamics,''
Nucl. Phys. B \textbf{454}, 185-206 (1995)
%doi:10.1016/0550-3213(95)00409-L
[arXiv:hep-th/9506035 [hep-th]].


%\cite{Schellstede:2016zue}
\bibitem{Schellstede:2016zue}
G.~O.~Schellstede, V.~Perlick and C.~L\"ammerzahl,
``On causality in nonlinear vacuum electrodynamics of the Pleba\'nski class,''
Annalen Phys. \textbf{528}, no.9-10, 738-749 (2016)
%doi:10.1002/andp.201600124
[arXiv:1604.02545 [gr-qc]].



%\cite{Bandos:2021rqy}
\bibitem{Bandos:2021rqy}
I.~Bandos, K.~Lechner, D.~Sorokin and P.~K.~Townsend,
``ModMax meets Susy,''
JHEP \textbf{10} (2021), 031
%doi:10.1007/JHEP10(2021)031
[arXiv:2106.07547 [hep-th]].





%\cite{Kuzenko:2000uh}
\bibitem{Kuzenko:2000uh}
S.~M.~Kuzenko and S.~Theisen,
``Nonlinear selfduality and supersymmetry,''
Fortsch. Phys. \textbf{49}, 273-309 (2001)
%doi:10.1002/1521-3978(200102)49:1/3\ensuremath{<}273::AID-PROP273\ensuremath{>}3.0.CO;2-0
[arXiv:hep-th/0007231 [hep-th]].


%\cite{Born:1938zve}
\bibitem{Born:1938zve}
M.~Born,
``A suggestion for unifying quantum theory and relativity,''
Proc. Roy. Soc. Lond. A \textbf{165}, no.921, 291-303 (1938)
%doi:10.1098/rspa.1938.0060

%\cite{Freidel:2013zga}
\bibitem{Freidel:2013zga}
L.~Freidel, R.~G.~Leigh and D.~Minic,
``Born Reciprocity in String Theory and the Nature of Spacetime,''
Phys. Lett. B \textbf{730}, 302-306 (2014)
%doi:10.1016/j.physletb.2014.01.067
[arXiv:1307.7080 [hep-th]].







\end{thebibliography}

\end{document}